\documentclass[11pt,letterpaper]{article}

\usepackage{dcolumn}
\usepackage{bm}

\usepackage{xcolor}

\usepackage{authblk}
\usepackage{doi}
\usepackage{graphicx}
\usepackage{amsmath}
\usepackage{amssymb}
\usepackage[margin=0.75in, letterpaper]{geometry}
\usepackage{cite}
\usepackage{hyperref}
\usepackage{url}
\usepackage{doi}

\newcommand{\leadauthor}[1]{\maketitle}
\newcommand{\significancestatement}[1]{}
\newcommand{\authorcontributions}[1]{\def\authorcontribs{#1}}
\newcommand{\correspondingauthor}[1]{}
\newcommand{\keywords}[1]{}
\newcommand{\dates}[1]{} 
\newcommand{\authordeclaration}[1]{} 
\newcommand{\dropcap}[1]{#1}
\newcommand{\matmethods}[1]{\section*{Materials and methods} \sloppy #1}
\newcommand{\showmatmethods}[1]{}
\newcommand{\acknow}[1]{
    \section*{Author contributions}\authorcontribs
    \section*{Acknowledgments}#1
}
\newcommand{\showacknow}[1]{}
\newcommand{\bibsplit}[1][1]{}
\newcommand{\supportinginformation}{
\newpage
\begin{center}
\LARGE Supporting Information 
\end{center}
\setcounter{figure}{0}
\renewcommand{\thefigure}{S\arabic{figure}}
}

\newcommand{\RNum}[1]{\uppercase\expandafter{\romannumeral #1\relax}}

\begin{document}

\title{Gliding microtubules exhibit tunable collective rotation driven by chiral active forces
}

\author[a]{Madhuvanthi Guruprasad Athani}
\author[a,b]{Nathan Prouse}
\author[c,d]{Niranjan Sarpangala}
\author[c,e]{Patrick Noerr}
\author[f]{Guillaume Schiano-Lomoriello}
\author[c]{Ankush Gargeshwari Kumar}
\author[c,g]{Fereshteh L. Memarian}
\author[f]{Jeremie Gaillard}
\author[f]{Laurent Blanchoin}
\author[c]{Linda S. Hirst}
\author[c]{Kinjal Dasbiswas}
\author[c]{Ajay Gopinathan}
\author[f,h,i]{Ond\v{r}ej Ku\v{c}era}
\author[a]{Daniel A.~Beller\footnote{To whom correspondence should be addressed. \makeatletter E-mail: d.a.beller@jhu.edu}}

\affil[a]{Department of Physics and Astronomy, Johns Hopkins University, Baltimore, MD 21218}
\affil[b]{Department of Physics and Astronomy, University of Southern California, Los Angeles, CA  90007}
\affil[c]{Department of Physics, University of California, Merced, CA 95343}
\affil[d]{Department of Physics and Astronomy, University of Pennsylvania, Philadelphia, PA 19104}
\affil[e]{Department of Pharmacology, University of California, San Diego, CA 92161}
\affil[f]{Laboratoire de Physiologie Cellulaire et V\'eg\'etale, Interdisciplinary
Research Institute of Grenoble, CEA/CNRS/Universit\'e Grenoble Alpes, 17 Avenue des
Martyrs, Grenoble, 38 054, France}
\affil[g]{Robert Bosch LLC, North American Research \& Technology Center, Sunnyvale, CA, 94085}
\affil[h]{Institute of Photonics and Electronics of the Czech Academy of Sciences, Chabersk\'a 57, 182 00 Prague, Czechia}
\affil[i]{Department of Engineering Technology, South East Technological University, Cork Road, Waterford, X91 K0EK, Ireland}


\leadauthor{Athani}

\significancestatement{
Chirality, the breaking of mirror symmetry, is ubiquitous in biology. However, much remains unknown about how chirality at the molecular scale leads to chirality in complex, collective behaviors of living systems at the cellular scale and beyond. We study this question using \textit{in vitro} experiments  on spontaneous collective rotations in traveling biological filaments, together with computational modeling. Our findings show that collisions between filaments play a central role in the collective rotations, due to a chiral bias in the direction of forces exerted by molecular motors. Our model suggests a route by which the features of large-collective motions can be tuned through molecular details in nature and in technology.
}

\authorcontributions{
D.A.B., K.D., and A.G.\ conceptualized the study; 
M.G.A. and D.A.B.\ designed the simulations; 
M.G.A., N.P., and P.N.\ performed the simulations and analyzed the computational results;
N.P.\ calculated bulk rotation predictions from pair collision data;
 L.B., O.K. and J.G.\ designed the experiments; 
G.S.-L., F.L.M., J.G., and O.K.\ performed the experiments;
N.S., A.G.K., and O.K.\ analyzed the experimental images;
L.H., K.D., A.G., L.B., O.K., and D.A.B.\ supervised the project;
M.G.A, O.K., and D.A.B. wrote the paper with input from all authors.}
\authordeclaration{The authors declare no competing interest.}
\correspondingauthor{\textsuperscript{1}To whom correspondence should be addressed. E-mail: d.a.beller\@jhu.edu}


\begin{abstract}
How chirality propagates across scales remains an open question in many biological and synthetic systems. An especially clear manifestation of this propagation is found in \textit{in vitro} gliding assays  of cytoskeletal filaments on surfaces, driven by molecular motors. These assays have become model systems of active matter dynamics, as they spontaneously organize into diverse dynamical states, including collective motions with chiral rotation. However, the microscopic mechanisms underlying these chiral collective dynamics have remained unclear. Here, we investigate rotating active nematic order in microtubule gliding assay experiments under two stabilization conditions, each on two types of substrates. We propose that chirality in active forces exerted by motors on microtubules represents a viable mechanism for this large-scale chirality. Using Brownian dynamics simulations of self-propelled, semiflexible filaments with chiral activity, we demonstrate that coherently rotating active nematic order  emerges by this mechanism even in the absence of curvature, i.e.\ shape chirality, of the constituent filaments. Moreover, we predict that the angular speed and  handedness of the collective rotation can be tuned by modulating filament stiffness. Our findings identify a new set of sufficient microscopic ingredients for predictable propagation of chiral handedness from the molecular to the material scale in living and active matter. 
\end{abstract}


\maketitle





\dropcap{C}hirality, the property of an object that is inequivalent to its mirror image even under any translation and rotation, is present in biological matter at all scales, from molecular building blocks \cite{Prelog1976, Bonner1995} to the systemic level \cite{hamada2002establishment, levin2005left}. Despite the crucial roles of chirality to life on many levels of organization, it remains unclear how molecular chirality leads to the macroscopic left-right asymmetries in multicellular organisms \cite{Tverdislov2013, Coutelis2014, Inaki2016}. At the cellular level, chirality is often associated with the cytoskeleton \cite{Vandenberg2013}, a network of protein polymers that play multiple roles in cell function, such as in motility, internal transport, and cell division. Numerous reports have explored links between the chiral structures of cytoskeletal filaments and the emergence of left-right asymmetries \cite{Hagmann1993, Wan2011, Chen2012, naganathan2016actomyosin, tee2015cellular, lebreton2018molecular, Neahring2024}. Similarly, experiments \textit{in vitro} with ensembles of cytoskeletal filaments and molecular motors have revealed signs of collective chiral behavior \cite{kim2018, Tanida2020, sciortino2023polarity}, though the underlying principles and broader implications remain largely unexplored.

Efforts to explain the emergence of chirality in cytoskeletal systems generally follow two lines of reasoning. First, the internal helical structure of cytoskeletal filaments can break left-right symmetry, either through torques arising during filament elongation \cite{middelkoop2021cyk} or through the action of molecular motors that recognize this structural asymmetry and move along helical paths, potentially generating additional torque in the process \cite{Ray1993, Nitzsche2008, Tee2015, Ramaiya2017, lebreton2018molecular, Meiner2024}. Second, the filaments themselves can be curved, and this curvature can induce chiral behavior, as demonstrated in the case of FtsZ (the prokaryotic homologue of tubulin) filaments \cite{Dunajova2023}.

One illustrative example of how filament chirality can give rise to chiral behavior in larger assemblies is provided by \textit{in vitro} experiments using microtubule–kinesin gliding assays. In these assays, microtubules—tubular cytoskeletal polymers composed of tubulin heterodimer subunits—are propelled across a substrate by kinesin motors anchored on the surface. When the density of microtubules is sufficiently high, they form long-range ordered (LRO) nematic states \cite{Memarian2021, Afroze2021, kim2018, Tanida2020, kuvcera2022actin, saito2017factors}. The director of this LRO nematic state is seen to rotate in a persistent counterclockwise (CCW) direction in some of these experiments \cite{kim2018, Tanida2020}.

With time represented as the $z$-axis, the rotating director field closely resembles that of a cholesteric (chiral nematic) ground state  (Fig.~\ref{fig:1}\textbf{G}). By analogy \cite{chen2025ambidextrousflocktwodimensionalchiral} with spontaneously ordered, time-periodic structures termed \textit{time crystals}, previously observed in certain quantum systems \cite{wilczek2012quantum}, this active state has been termed \textit{time cholesteric} \cite{maitra2024activity}.

Although the formation and stability of long-ranged nematic order in these gliding assays have been thoroughly investigated \cite{athani2023symmetry, Tanida2020, saito2017factors}, the origin of the nematic director rotation remains uncertain. Some studies have found evidence that the large-scale rotation is not a collective effect at all, but rather a synchronized version of the individual rotation that each filament would perform in isolation as it follows a circular trajectory \cite{Tanida2020}. However, we find in this work that the collective rotation rate can differ from that predicted by measured filament curvatures by orders of magnitude. More specifically, it has been proposed that the circular trajectories of individual filaments are determined by a signed, nonzero intrinsic curvature of the equilibrium filament contour \cite{kim2018}. This hypothesis is challenged by the observation that individual microtubules turn clockwise (CW) in gliding assays with dynein motors \cite{Sumino2012} as opposed to  counterclockwise with kinesin motors \cite{kim2018}, pointing to activity rather than equilibrium shape as essential to chiral motion. It has also been proposed that collective rotations of gliding microtubules require chiral pair-interactions, specifically a collision-induced torque \cite{Hiraiwa2022}, rather than merely individual-filament properties. 

Thus, consensus has not yet emerged on the fundamental mechanisms by which molecular chirality affects collective dynamics in gliding cytoskeletal filaments. Does microtubule rotation arise simply from microtubule curvature, or do both rotation and curvature result from the chiral action of motors? By what route does the helicity of the microtubule's molecular structure connect to its chiral self-propulsion, mechanical response to collisions, and collective rotation?  

In this work, we demonstrate that chirality in active forces at the molecular scale can produce macroscale time cholesteric behavior, without requiring shape chirality. In microtubule gliding assays, chirality may enter into the active forces exerted by motor proteins through the helical supertwist of the protofilaments along which the motors walk, even if the microtubule shape displays no chiral bias \cite{Ramaiya2017,Meiner2024}. To test this hypothesis, we combine experimental investigation using microtubule-kinesin gliding assays with Brownian dynamics simulations to explore the emergence and properties of the time cholesteric state. Our results show that the structural and mechanical properties of the microtubules can tune the amplitude, stability, and -- strikingly -- even the direction of this collective rotation. Chiral self-propulsion, directed at a fixed angle relative to the filament contour, offers a viable mechanism for experimentally observed collective rotations and predicts a wealth of new phenomena in other parameter regimes. Our results hold implications for understanding the role of molecular motors in the propagation of chirality across scales in biology, and may inform the design of chiral active matter more broadly \cite{tan2022odd, liebchen2017collective}.

\section*{Collective rotation emerges from interactions, not filament curvature}

\begin{figure*}[t]
    \includegraphics[width=17.8cm]{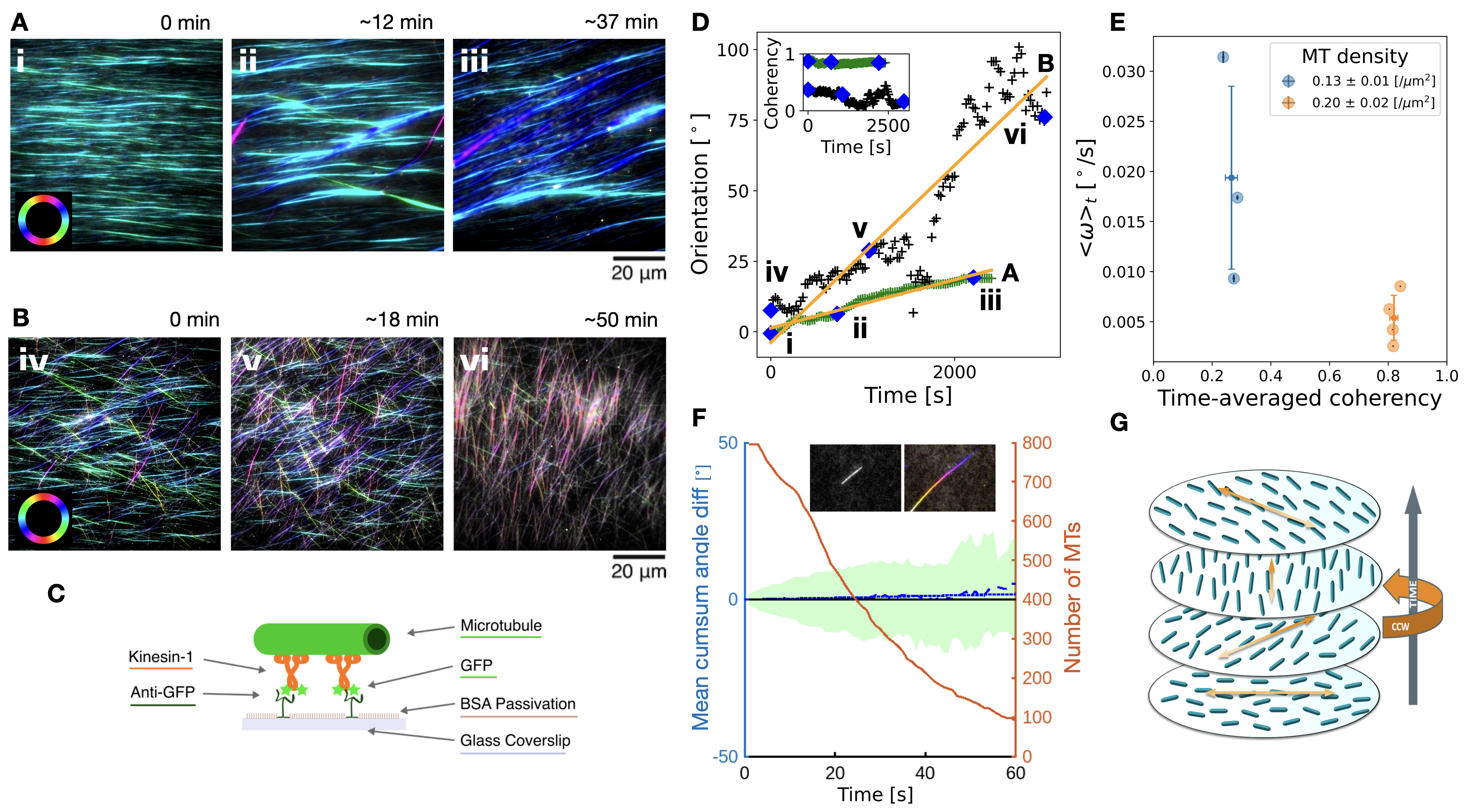}
    \caption{Microtubule gliding assay self-organizes into rotating nematic state. \textbf{A, B.} Fluorescence micrograph time-series of  microtubule motility assays, color-coded by the local orientation of microtubules. The microtubule surface density in filaments per square micrometer is $0.20\pm 0.02$ in \textbf{A} and $0.13\pm 0.01$ in \textbf{B}. \textbf{C.} Schematic representation of the experimental system. \textbf{D.} The temporal profile of the nematic director orientation, and of the coherency (inset), for the experiments imaged in \textbf{A,B}. \textbf{E.} Relation between the coherency of the nematic order and the rotation rate of the nematic director. Data from individual experiments are represented as error bars (mostly smaller than filled-circle markers). For each of two microtubule surface densities tested, whisker plots indicate interval of $\pm 1$ standard deviation about the median along both axes. \textbf{F.} Cumulative angle difference of individual gliding microtubules as a function of time. The blue data points represent the median at each time $t$ since the beginning of the measurement on the respective microtubule, and the green shading represents the interquartile range. The blue dashed line is a linear fit weighted by the number of contributing microtubules at each time, which decays as plotted in orange. The inset images show a typical fluorescence micrograph of a microtubule in these experiments (left) and the corresponding maximum intensity projection of the consecutive frames showing the gliding motion (right, temporal color-coded). \textbf{G.} Schematic representation of the time-cholesteric behavior.}
    \label{fig:1}
\end{figure*}

To examine  rotating active nematic order in a microtubule gliding assay driven by kinesin motors as reported in the literature \cite{kim2018, Tanida2020}, we immobilized truncated kinesin-I motors onto the surface of an imaging glass chamber at an approximate density of 17.4 molecules per square micrometer (Fig.~\ref{fig:1}\textbf{C} and Materials and Methods, following the methodology of Ref.~\cite{Kucera2022Actin}). 

We then introduced GMPCPP-microtubules, preassembled to a median length of 
{
3.35 $\pm$ 0.23} $\mu$m, into the chamber. Using TIRF microscopy, we observed the microtubules binding to the surface-attached kinesin motors. Upon the addition of adenosine 5’-triphosphate (ATP), the kinesin motors propelled the microtubules, resulting in their gliding motion along the surface.

Previous studies have demonstrated that both the degree and stability of nematic ordering in gliding microtubules correlate positively with microtubule surface density and the concentration of crowding agents \cite{saito2017factors}. Accordingly, we optimized the conditions of our assay (see Methods) and employed a buffer containing 0.327\% (wt/vol) of 63-kDa methylcellulose. Under these conditions, microtubules at a surface density of approximately 0.20 $\pm$ 0.02 filaments per square micrometer collectively assembled into a state with long-range nematic order (Fig.~\ref{fig:1}\textbf{A} and \textit{SI Movie S1}). To quantify the coherence of this nematic ordering, we measured the nematic order parameter (Materials and Methods), obtaining a value of 
{
0.82 $\pm$ 0.01 ($n = 4$ experiments)}, where $1$ represents perfect order. This order remained stable over time throughout the experiment (Fig.~\ref{fig:1}\textbf{D} inset). The nematic director exhibited a consistent CCW rotation at an average angular velocity of 
{
0.005 $\pm$ 0.002 degrees per second ($n = 4$ experiments}; Fig.~\ref{fig:1}\textbf{E}), in qualitative agreement with prior reports \cite{kim2018, Tanida2020}. We note that doubling the surface density of kinesin-I motors to approximately $34.8$ molecules per square micrometer resulted in inconsistent formation of active nematic order (SI Fig.~S2).

To examine the behaviors of individual filaments, we decreased microtubule surface density to approximately 0.0037 $\pm$ 0.0014 filaments per square micrometer, enabling extended observation of individual filaments. We tracked these isolated microtubules and removed track segments influenced by collisions (Materials and Methods). To evaluate whether microtubules exhibited a directional turning bias, we plotted the cumulative angular deviation relative to each filament’s initial orientation as a function of time and traveled distance. We found that angular distributions were broad (Fig.~\ref{fig:1}\textbf{F}), nevertheless, a slight yet statistically significant preference emerged for CCW turning ($p = 5.75 \times 10^{-11}$, Mann–Kendall test, $n = 796$ tracks from 30 experiments), with an average rotation rate of 0.029 degrees per second (95\% confidence bounds of 0.0225 and 0.03472), corresponding to 0.1227 degrees per $\mu$m traveled (SI Fig.~S1). Individual-microtubule rotation rates are much faster, by nearly an order of magnitude, than the measured collective rotation rate.

We also analyzed the signed curvatures of microtubules in these experiments to evaluate whether their  shapes could account for the turning observed during gliding. From the measured curvature and the average gliding velocity, we calculated the angular rotation rate that would result if a microtubule were to follow a trajectory defined by its shape. This yielded a value of approximately \(1.74 \times 10^{-7}\) degrees per second. As this is several orders of magnitude lower than the turning rates observed experimentally,  our data are inconsistent with a model in which collective rotation reflects circular filament trajectories matching the average curvature of filament contours. 

Furthermore, if the rotation of the nematic director in the collective experiments were driven solely by the curvature of individual microtubule trajectories, we would expect collisions between filaments to disrupt that individually preferred rotation. Collisions will be less frequent when the nematic order is stronger, so we would expect the degree of nematic order to correlate positively with the observed rotation rate.
To test for a correlation between nematic order and collective rotation rate, we decreased the microtubule density in the collective dynamics experiment to $0.13 \pm 0.01$ filaments per square micrometer, holding all other assay parameters constant (Fig.~\ref{fig:1}\textbf{B} and \textit{SI Movie S2}). This change reduced the degree of nematic order—yielding a nematic order parameter of 
{
$0.27 \pm 0.02$} ($n = 3$ experiments). Strikingly, the rotation rate increased to 
{
$0.019 \pm 0.009$} degrees per second (Fig.~\ref{fig:1}\textbf{D},\textbf{E}). Because collisions between microtubules are expected to be more frequent when nematic order is lower, the associated increase in rotation rate suggests that such collisions play a significant role in driving the collective rotation behavior.




\section*{Modeling supports chiral self-propulsion as driver of rotation, via collisions}
\begin{figure*}[t]
    \includegraphics[width=17.8cm]{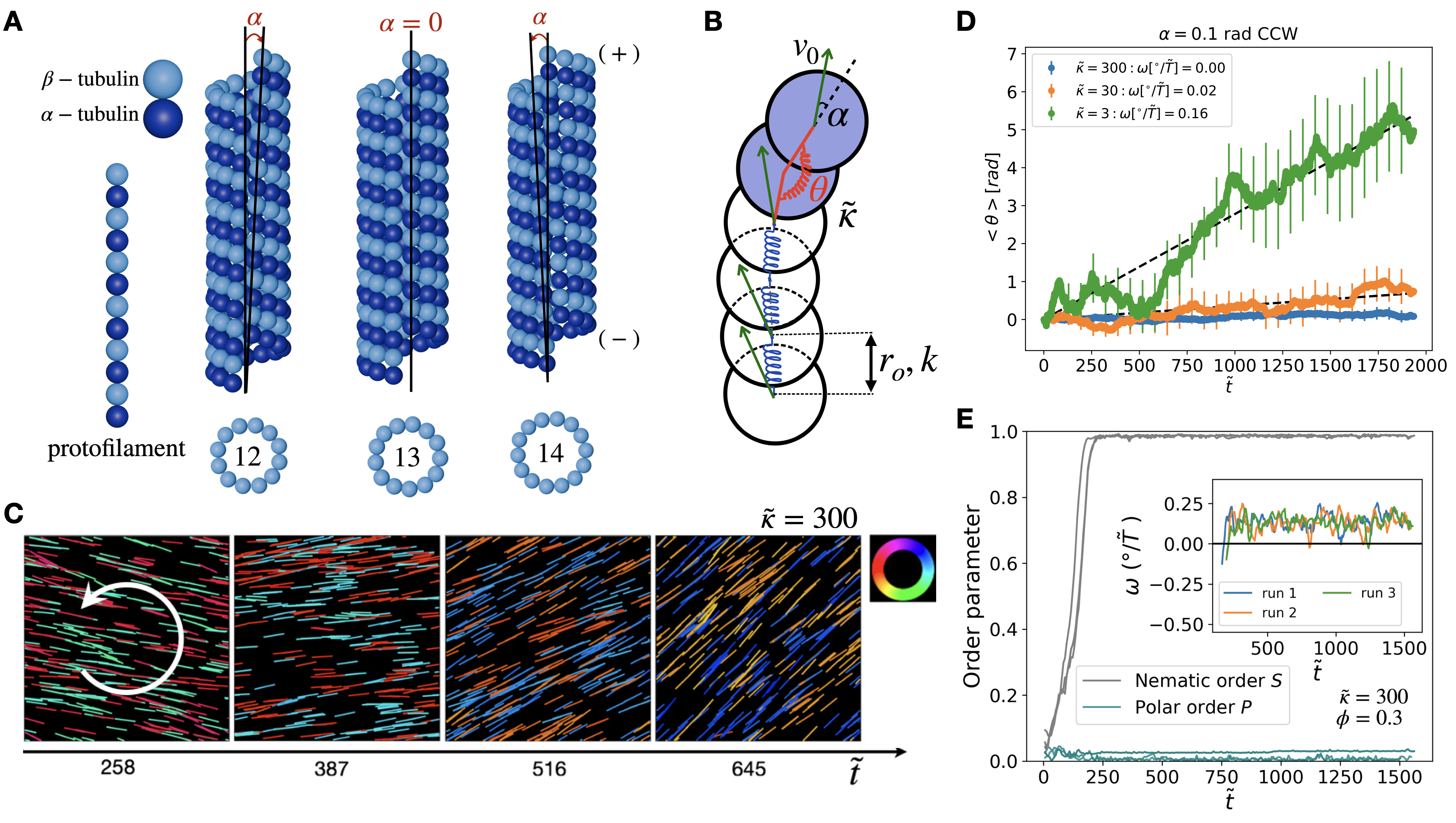}
    \caption{\textbf{A.} Illustration of microtubule lattice structures with $N = 12$, $13$ and $14$ protofilaments. All examples shown have a 3-start helix. The skew angle $\alpha$ of the protofilaments from the tube axis is noted, with different signs for $N=12$, $14$ reflecting differing handedness of supertwist. \textbf{B.} Schematic illustration of the bead-spring chain model of a microtubule, with Hookean linear springs (blue coils) of rest length $r_0$ and spring constant $k$ between neighboring beads, harmonic bending rigidity $\tilde{\kappa}$ (red coil), and active force (green arrows), with corresponding active speed $v_0$, oriented at a signed skew angle $\alpha$ relative to the local tangent. \textbf{C.} Simulation time series snapshots showing CCW rotation of nematic director for bending rigidity $\tilde{\kappa} = 300$, skew angle $\alpha = 0.1$rad CCW, area fraction $\phi = 0.3$ and with time $\tilde t$ scaled in units of $\tilde{T}=l/v_0$ where $l$ is the filament length. The color wheel legend, here and in subsequent figures, indicates the orientation of each filament, averaged over the tangents at each bead. \textbf{D.} Average change in orientation of an ensemble ($n = 10$) of single filaments with time for active force skew angle $\alpha = 0.1$rad. Rotation rates obtained from a linear fit indicate no significant rotation for high bending rigidities $\tilde \kappa$. \textbf{E.} Time evolution of global nematic and polar order parameters, overlaying data from the simulation imaged in \textbf{C} and two other runs with the same parameters, all with randomized isotropic initial conditions. Inset shows the time-evolution of the angular velocity $\omega$ of the nematic director, plotted in units of $^{\circ}/\tilde{T}$. $\omega$ is positive, indicating CCW rotation.}
    \label{fig:2}
\end{figure*}

Having demonstrated that time cholesteric behavior cannot be fully explained by intrinsic curvature in individual microtubule trajectories, we turned to computer simulations to clearly distinguish the contributions of filament collisions from those of single-filament turning to the nematic director rotation. Our model demonstrates the viability of chiral active forces, in the absence of shape chirality, as a mechanism for time cholesteric behavior. 

Our focus on chiral active forces is motivated by the broken mirror symmetry found in the internal structure of microtubules. Each microtubule consists of tubulin heterodimers organized into parallel protofilaments that form a tubular structure, typically composed of $N=12$ to $15$ protofilaments (Fig.~\ref{fig:2}\textbf{A}). This assembly creates a rhombic lattice of tubulin proteins in the microtubule wall. Protofilaments either align parallel to the tube axis ($N=13$) or adopt a long-pitch helical twist—right-handed for $N<13$ and left-handed for $N>13$—resulting in a characteristic skew angle $\alpha$ relative to the tube axis \cite{Pampaloni2008, Nakamura2020, Zhang2018, Amos2007}. Kinesin-1 motors move directionally towards the microtubule plus-end along individual protofilaments \cite{Ray1993, Nitzsche2008}. In gliding assays, surface-bound kinesin motors propel microtubules with their minus-end oriented forwards. When $N \neq 13$, the helical protofilament arrangement creates a slight directional mismatch between the local microtubule tangent and the motor-induced propulsion force.

We modeled the microtubule-kinesin gliding assay as a dry active system using a two-dimensional Brownian dynamics simulation. Microtubules were represented as self-propelled, semiflexible filaments, following the approach of Ref.~\cite{athani2023symmetry}, with a modification of the active force to introduce chirality. Each semiflexible filament consisted of beads connected by Hookean springs with spring constant $k$ and rest length $r_0$. We keep the number of beads in each filament fixed at $30$, giving the filament a nominal length $\ell = 31r_0$. Bending rigidity $\tilde{\kappa}$ acts between three consecutive beads when the angle $\theta$ between them deviates from a straight configuration through harmonic bending energy (Fig.~\ref{fig:2}\textbf{B}). 

 We model the chiral active forces as acting with constant magnitude and oriented at a small angle $\alpha$ to the local tangent at each bead. The active force produces a self-propulsion speed $v_0$ in the absence of other forces and acts on every bead at each time step.  For GMPCPP microtubules as in Fig.~\ref{fig:1}, we expect an excess of $N=14$ microtubules over $N=12$, so we model this system with a positive $\alpha$ corresponding to a typically left-handed helicity of supertwist \cite{Meiner2024}.  For simplicity, we take $\alpha$ to be the same value for all filaments in a given simulation. As $\alpha$ is the only chiral term in our system, all of our findings with $\alpha > 0$ will hold equally for $\alpha<0$ after inversion. 


The interactions between the center of beads $i$ and $j$ of different filaments are modeled as a Weeks-Chandler-Andersen (WCA) potential  $U_{\mathrm{WCA}}$ \cite{Weeks1971} shifted to be finite at zero separation ($r_{ij} = 0$) using a parameter $r_{\mathrm{shift}} \neq 0$: 
\begin{align}
    U_{\mathrm{WCA}}\left(r_{ij}\right)   
    &= \begin{cases}
            4 \varepsilon\left[
                \left(
                    \frac{\sigma}{r_{ij}+r_{\mathrm{shift}}}
                \right)^{12} - \left(
                    \frac{\sigma}{r_{ij}+r_{\mathrm{shift}}}
                \right)^{6}
            \right]+\varepsilon, 
            & 
            r_{ij} \leq 2^{1 / 6} \sigma 
            \\ 0, 
            & r_{ij}>2^{1 / 6} \sigma,
    \end{cases} 
    \label{wca_eq}
\end{align}
This shift models the observation that colliding microtubules frequently delaminate from the surface and cross over each other. 
The formation of LRO states requires that active filaments be allowed to interpenetrate at a modest energy cost $\varepsilon$  in the absence of explicit aligning or attractive terms in the interaction energy \cite{moore2020collective, athani2023symmetry}. We exclude the latter types of interactions because they model the bundling effects of depletants, but recent microtubule-kinesin gliding assay experiments on lipid substrates achieve LRO states without the use of depletants \cite{Memarian2021}; we therefore omit explicitly aligning or attractive terms from the filament interactions in our model, which mimic the bundling effect of depletants.  

Throughout this work, we use the parameters $r_{\mathrm{shift}} = 0.1$ and $\varepsilon = 10^{-8}$ that were found in Ref.~\cite{athani2023symmetry} to produce LRO nematic states in the $\alpha=0$ achiral case. Simulation length $\tilde x$ is scaled in the units of the range of the WCA repulsion potential $\sigma$ as $\tilde x  = 2^{1/6}\sigma$. The Hookean spring rest length is $r_0 = 0.5 \tilde x$ in these units. We employ periodic boundary conditions in a simulation box whose size is $L_x = 150$ except where stated otherwise.

With the introduction of chiral activity through the small angle $\alpha$ between the active force and the local tangent, we characterized the chirality of the emergent LRO states. Beginning with $\alpha = 0.1$rad CCW, we found a time-cholesteric state, characterized by global  nematic order with a nematic director that steadily rotates with a consistent handedness (Fig.~\ref{fig:2}\textbf{C} and \textit{SI Movie S3}). It was shown in Ref.~\cite{athani2023symmetry}, from which this simulation model is adapted, that a nondimensionalized bending rigidity of $\tilde{\kappa} \approx 250$ characterizes typical microtubules in the gliding assay experiments of Ref.~\cite{Memarian2021}. Here, using a similar bending rigidity $\tilde \kappa = 300$, we find that chiral activity causes the director to rotate CCW, i.e.\ with the same handedness as the active self-propulsion relative to the local tangent. This finding confirms that a chiral active force, in the form of off-tangent self-propulsion, represents a viable source of collective rotation.

It is not obvious \textit{a priori} from our model whether chiral activity will cause isolated filaments to rotate with any preferred handedness; a perfectly rigid ``active screw'' could translate at angle $\alpha$ to its tangent without rotation of that tangent \cite{banerjee2024active}.  Simulations of single filaments with $\alpha = 0.1\,\textrm{rad}$ show that, while more flexible filaments preferentially rotate with the same handedness as $\alpha$, filaments with higher $\tilde \kappa$ corresponding to the microtubules in our experiments do not exhibit any significant average rotation (Fig.~\ref{fig:2}\textbf{D}). We find no overlap of the high-$\tilde \kappa$ regime in which LRO states emerge \cite{athani2023symmetry} and the low-$\tilde \kappa$ regime in which individual filaments rotate with a consistent handedness. This feature of our model allows us to clearly establish that time cholesterics can exist in parameter regimes where individual filaments do not rotate. Filament interactions, rather than individual behavior, are thus implicated as the route by which chirality propagates to the global scale from the active units.


\def\tunableSectionName{Collective rotation is tunable through bending rigidity} 
\section*{\tunableSectionName}\label{sec:tunable}
 
\begin{figure*}[t]
    \includegraphics[width=17.8cm]{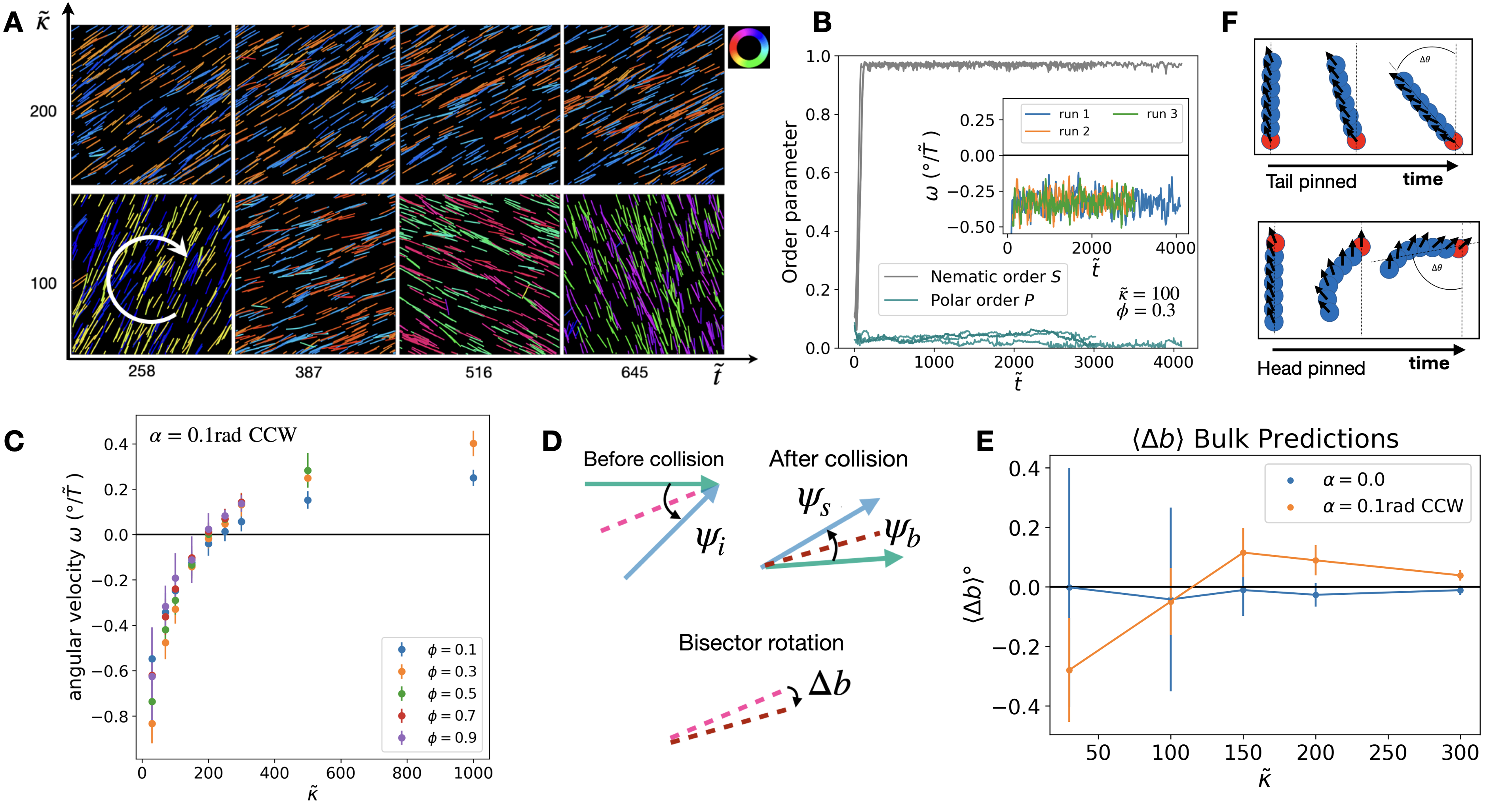}
    \caption{\textbf{A.} Simulation snapshots showing that the direction of the nematic director's collective rotation depends on $\tilde{\kappa}$. The two rows are time series from two simulations, with time $\tilde t$ measured in units of $\tilde T$ and with $\alpha = 0.1$ rad CCW, area fraction $\phi = 0.3$. \textbf{B.} Time evolution of global nematic and polar order parameters for $\tilde{\kappa} = 100$, overlaying data from a 3-run ensemble with randomized isotropic initial conditions. Insets show the time-evolution of the angular velocity $\omega$ of the nematic director;  consistently negative values indicate CW rotation. \textbf{C.} Angular velocity of the LRO nematic director as a function of bending rigidity $\tilde{\kappa}$ for varying area fraction $\phi$, with $\alpha = 0.1$rad CCW. Each data point is the mean of a 3-run ensemble; error bars show standard deviation. \textbf{D.} Illustration of a pair collision event and the associated angles measured in the simulation: the incident and scattered relative angles $\psi_i$ and $\psi_s$, and the change $\Delta b$ in the bisector orientation  $\psi_b$. \textbf{E.} Expected rotation per collision of the filament pair bisector $\Delta b$ as calculated from pair collision simulations by weighting $\Delta b$ for each incident collision angle by the estimated rate of collisions at that  angle in the bulk steady-state. One achiral and one chiral case are plotted. Error bars show the standard deviation.}
    \label{fig:3}
\end{figure*}

In the prior application of our simulation model to the achiral ($\alpha=0$) case, the bending rigidity $\tilde \kappa$ of filaments was found to have a dramatic impact on the LRO state: larger $\tilde \kappa$ significantly delays the development of polar order, leading to long-lived transient active nematic states \cite{athani2023symmetry}. Therefore, we explore here the effects of varying $\tilde \kappa$ on the time cholesteric state with $\alpha\neq 0$. For simplicity, our model assumes that $\tilde \kappa$ has the same value for all microtubules at all times in a given simulation, although for real microtubules such mechanical properties are closely linked to the tubulin structure, which can be heterogeneous \cite{motta2023beyond}.



So far, we have demonstrated that $\tilde \kappa=300$ leads to collective rotation with the same handedness as the chiral self-propulsion. Surprisingly, when we reduced the bending rigidity to $\tilde{\kappa} < 200$ while keeping the bead area fraction constant at $\phi = 0.3$, the collective rotation reversed direction, becoming CW (Fig.~\ref{fig:3}\textbf{A} bottom row and \textit{SI Movie S4}). It is important to emphasize that this reversal in macroscopic handedness occurs even though the skew angle of the active force was held fixed at $\alpha = 0.1$ rad. The transition is continuous, as the collective rotation vanishes at $\tilde{\kappa} \approx 200$, producing an active nematic with constant director  (Fig.~\ref{fig:3}\textbf{A} top row and \textit{SI Movie S4}). These behaviors were consistent over three-run ensembles at each $\tilde \kappa$, as shown in the inset to Fig.~\ref{fig:3}\textbf{B} for $\tilde \kappa = 100$. The dependence of angular velocity $\omega$ (time- and ensemble-averaged) on bending rigidity $\tilde \kappa$ is summarized in Fig.~\ref{fig:3}\textbf{C}, demonstrating that tuning $\tilde{\kappa}$ allows control over the sign and magnitude of $\omega$. 

To examine the symmetry of the emergent long-range order, we measure the global nematic and global polar order parameters (Materials and Methods). Starting from random initial conditions, we find strong nematic order, and an absence of polar order, in both the CW and CCW collective rotation regimes (Fig.~\ref{fig:3}\textbf{B} and Fig.~\ref{fig:2}\textbf{E} respectively). This stable nematic order is in stark contrast with the gradual decay of the LRO nematic transient state into the LRO polar steady state observed in achiral filaments in Ref.~\cite{athani2023symmetry}. It appears likely that the collective rotation interferes with the stochastic sorting of filaments into an increasing majority direction and an oppositely oriented minority direction, which underlies the rise of polar order in the achiral system. On this point, we note that angular velocities on the order of $0.1 ^\circ / \tilde T$, as we measure here, imply a rotation period comparable to or smaller than the timescales for saturation of polar order among achiral filaments as found in Ref.~\cite{athani2023symmetry}.


To better understand how interactions between chiral filaments lead to net rotation, we ran a large number of controlled ``scattering'' simulations, each containing just two filaments (SI Section II). The results of these simulations show a clear tendency toward  alignment between colliding filaments. We measure the net rotation resulting from  each collision through the rotation $\Delta b$ of the bisector between the orientations of the two filaments, calculated through their contour-averaged tangent vectors (Fig.~\ref{fig:3}\textbf{D}). Achiral filaments necessarily have $\left<\Delta b\right> = 0$. In contrast, when $\alpha = 0.1$ rad, a careful extrapolation from our pair-collision data predicts that $\Delta b$ is positive at large $\tilde \kappa$ and negative at small $\tilde \kappa$ (Fig.~\ref{fig:3}\textbf{E}). This is qualitatively consistent with our findings in the bulk, suggesting that collective rotation in both handedness regimes arises from filament collisions. The iterative scheme used to obtain these predictions is described in SI Section II, with explicit consideration for the dependence of $\Delta b$ on pre-collision angle as well as the relative frequencies of collisions at each such angle (SI Fig.~S3-S7). Thus, a collision-induced torque akin to that assumed by the model of Ref.~\cite{Hiraiwa2022} emerges from our model, without breaking mirror symmetry in the filament interaction rules. 

An intuitive explanation for the dependence of collision-induced rotation on filament bending rigidity is revealed when we consider two extreme scenarios for hindrance of a single filament's self-propulsion: We imagine pinning in place either the ``head'' (foremost bead) or the ``tail'' (aftmost bead) of the bead-spring chain. As shown in Fig.~\ref{fig:3}\textbf{F}, for a filament with active force direction oriented CCW from the forward tangent, the tail-pinned filament clearly rotates CCW whereas the head-pinned filament rotates CW. In a typical collision, the head of the impinging filament initiates either a crossover, by passing through the second filament, or an aligning event that rotates the two filaments into parallel or antiparallel orientations \cite{Huber2018, Tanida2020}. The overall rotation produced by an aligning event is likelier to resemble that of head-pinning (CW) than tail-pinning (CCW), as the impinging filament is hindered at its head. Furthermore, a smaller bending rigidity increases the likelihood of alignment as opposed to crossover (SI Fig.~S3{\bf B},{\bf C}), so we expect low $\tilde \kappa$ to be associated with a CW bias. In contrast, high $\tilde \kappa$ favors crossovers because of the higher energy required cost to deflect the impinging filament. During a crossover, both filaments impede each others' active motion until the tail of one leaves contact with the other, increasing the likelihood that the collision's overall rotation handedness matches that of pure tail-pinning (CCW). These arguments provide qualitative insight into the dependence of rotation handedness on flexibility, though we caution that quantitative details of  Fig.~S3{\bf C},{\bf E} are also sensitive to more subtle considerations, such as the dependence of collision duration on both filament flexibility and incident angle.


\section*{Experimental evidence for collision-driven, rigidity-modulated collective rotation}

\begin{figure*}[t]
    \includegraphics[width=17.8cm]{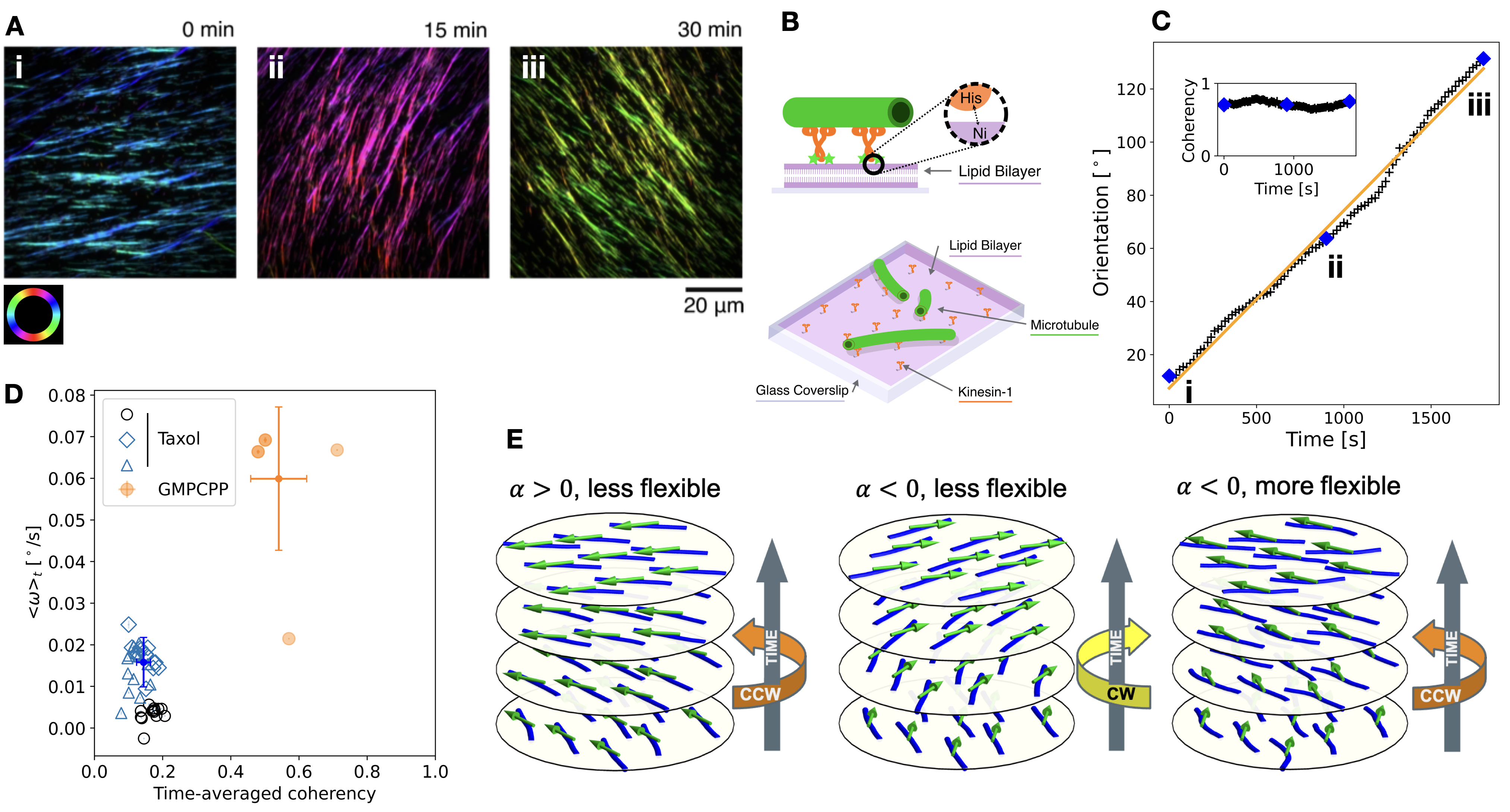}
    \caption{Microtubule motility gliding assay system on lipid bilayer self-organizes into rotating nematic state. \textbf{A.} Fluorescence micrographs of the GMPCPP-microtubule motility assay on lipid bilayer, color-coded by the local orientation of microtubules. \textbf{B.} Schematic representation of the assay.\textbf{C.} The temporal profile of the nematic director orientation (for data presented in \textbf{A}) and the coherency (inset). \textbf{D.} Relation between the coherency of the nematic order and the rotation rate of the nematic director. Coherency is calculated over a square field of view with side length $81.9$ $\mu$m. For GMPCPP microtubules, each data point comes from a separate experiment. For taxol-stabilized microtubules, data are presented from multiple fields of view in each of three experiments, indicated by three marker shapes with blue marker shapes representing experiments with MT density $\approx 0.45 \text{ filaments}/\mu \mathrm{m}^2$ and black circles with MT density $\approx 0.95 \text{ filaments}/\mu \mathrm{m}^2$. The standard deviation associated with each measurement is plotted as error bars within each marker, generally too small to see. For each of the two microtubule types, the median and one standard deviation along each axis are represented by the large whiskers. For taxol-stabilized MTs, median and standard deviation shown in large whiskers was computed using the data points shown in blue. \textbf{E.} Schematic illustration of the proposed double reversal of collective rotation handedness: Changing the sign of $\alpha$ in rigid microtubules replaces CCW with CW rotation, but then decreasing $\tilde \kappa$ restores CCW rotation.}
    \label{fig:4}
\end{figure*}

We have identified bending rigidity as a key parameter influencing both the rate and direction of the emergent time-cholesteric behavior in our model. To experimentally test this prediction, we turned to gliding assays using taxol-stabilized microtubules, which have a lower bending modulus compared to the previously used GMPCPP-microtubules \cite{Hawkins2013}. 

We initially used a glass substrate, as we did for GMPCPP microtubules, in our experiments examining collective effects of taxol-stabilized microtubules (average length $5.8 \pm 0.8\, \mu\text{m}$, average density $0.19 \pm 0.06$ filaments per square micrometer). While these experiments exhibited  nematic order (nematic order parameter = 0.633 $\pm$ 0.12), the microtubules formed pronounced nematic lanes \cite{Memarian2021}, which complicates the comparison with the uniform-density time cholesteric picture. The time-averaged angular speed $\omega$ of the nematic director was measurable but was inconsistent in sign across an ensemble of $n=6$ experiments, giving an ensemble average of $\omega \approx 0.0005 \pm 0.011$ degrees per second whose standard deviation is larger than its mean (SI Fig.~S8). 
The increased bundling and lane formation in taxol microtubules likely indicates an enhancement in filament pair alignment rather than crossover upon collision \cite{Sumino2012,  Huber2018, athani2023symmetry}.

To minimize density variations in favor of a synchronized director rotation through the field of view, we conducted additional experiments under conditions found previously to promote active nematic states of approximately uniform density \cite{Memarian2021}: 
%
We transferred our gliding assay onto a supported lipid bilayer, to which we added motors with a density comparable to that in experiments on glass (Fig.~\ref{fig:4}\textbf{B}; see Methods and Supplementary Information). Using fluorescence recovery after photobleaching (FRAP) assays, we confirmed that the motors were free to diffuse laterally (SI Fig.~S12). 

Under these conditions, we observed the formation of stable nematic order for both GMPCPP-microtubules (Fig.~\ref{fig:4}\textbf{A} and SI Movie S6) and taxol-stabilized microtubules. This ordering was coherent over a large area (SI Fig.~S9) and caused by kinesin motor activity, as confirmed by control experiments, and not by hydrodynamic flow during buffer exchange or microtubule loading (SI Fig.~S9). The mean ordering parameter was %
{
$0.54 \pm  0.08$ for GMPCPP-microtubules ($n = 6$ experiments}; average filament surface density = $0.24 \pm 0.05$ filaments per square micrometer), and {$0.14 \pm 0.02$ for taxol-stabilized microtubules ($n = 2$ experiments; average filament surface density $\approx 0.45$ filaments per square micrometer)}. In both filament types, the nematic order remained stable (albeit quite low for taxol-stabilized microtubules), and time cholesteric behavior was evident with director rotations that were coherent, spatially homogeneous, and approximately constant over time (Fig.~\ref{fig:4}\textbf{C}; SI Fig.~S10, S11). 

For both GMPCPP-microtubules and taxol-stabilized microtubules, the nematic director consistently rotated in a CCW direction, with an average angular velocity of 
{
$0.06 \pm 0.08$ degrees per second for GMPCPP-microtubules ($n = 6$ experiments), and $0.016 \pm 0.006$ degrees per second for taxol-stabilized microtubules (Fig.~\ref{fig:4}\textbf{D})}. The observed differences between the two types of microtubules were systematic and are unlikely to have resulted from random variation ($p = 0.007$, T-test for the means of two independent samples of scores, $scipy.stats.ttest\_ind$). The low nematic order among taxol-stabilized microtubules demands caution in interpreting the results shown in Fig.~\ref{fig:4}\textbf{D}. Nonetheless, the data taken over multiple experiments and many fields of view consistently indicate that taxol-stabilized microtubules collectively rotate CCW, with a slower angular speed than GMPCPP microtubules. 

The observation that taxol-stabilized microtubules collectively rotate CCW, like the GMPCPP microtubules, is remarkable in light of their structural differences: Whereas GMPCPP microtubules are expected to have an excess of 14-protofilament microtubules and thus average $\alpha > 0$, taxol-stabilized microtubules are expected to have an excess of 12-protofilament microtubules and thus average $\alpha < 0$ \cite{kim2018}. Thus, we have found that taxol-stabilized microtubules collectively rotate with the opposite handedness to the chiral active force generated by kinesin motor stepping along protofilaments. This is consistent with our model if we identify taxol-stabilized microtubules as belonging to the low-bending-rigidity regime of Fig.~\ref{fig:3}\textbf{C}. There, with $\alpha = 0.1$ rad, we observed CW collective rotation for $\tilde \kappa \lesssim 200$; this result directly implies that with $\alpha = -0.1$ rad, CCW rotation is predicted for the same $\tilde \kappa$ range. While it is challenging to give a precise numerical value for bending rigidity in this active system, the approximate value $\tilde \kappa = 250$ calculated in Ref.~\cite{athani2023symmetry} for the experiments of Ref.~\cite{Memarian2021} is notably close to the threshold value $\tilde \kappa \approx 200$ found in this work, so it is plausible that taxol-stabilized microtubules belong to the low-$\tilde \kappa$ regime. In contrast, the more rigid GMPCPP microtubules are likely to belong to the high-$\tilde \kappa$ regime in which the collective rotation the same handedness as the active force. Thus, the fact that both types of microtubules rotate CCW (Fig.~\ref{fig:4}\textbf{D}) on lipid substrates could result from two canceling ``sign flips'' of $\omega$ for the taxol case: one due to $\alpha < 0$, the other due to low $\tilde \kappa$. A schematic of this logic is presented in Fig.~\ref{fig:4}{\textbf{E}}.


\section*{Modeling predicts regimes of dynamical bistability and inhomogeneity}

\begin{figure*}[t]
    \includegraphics[width=17.8cm]{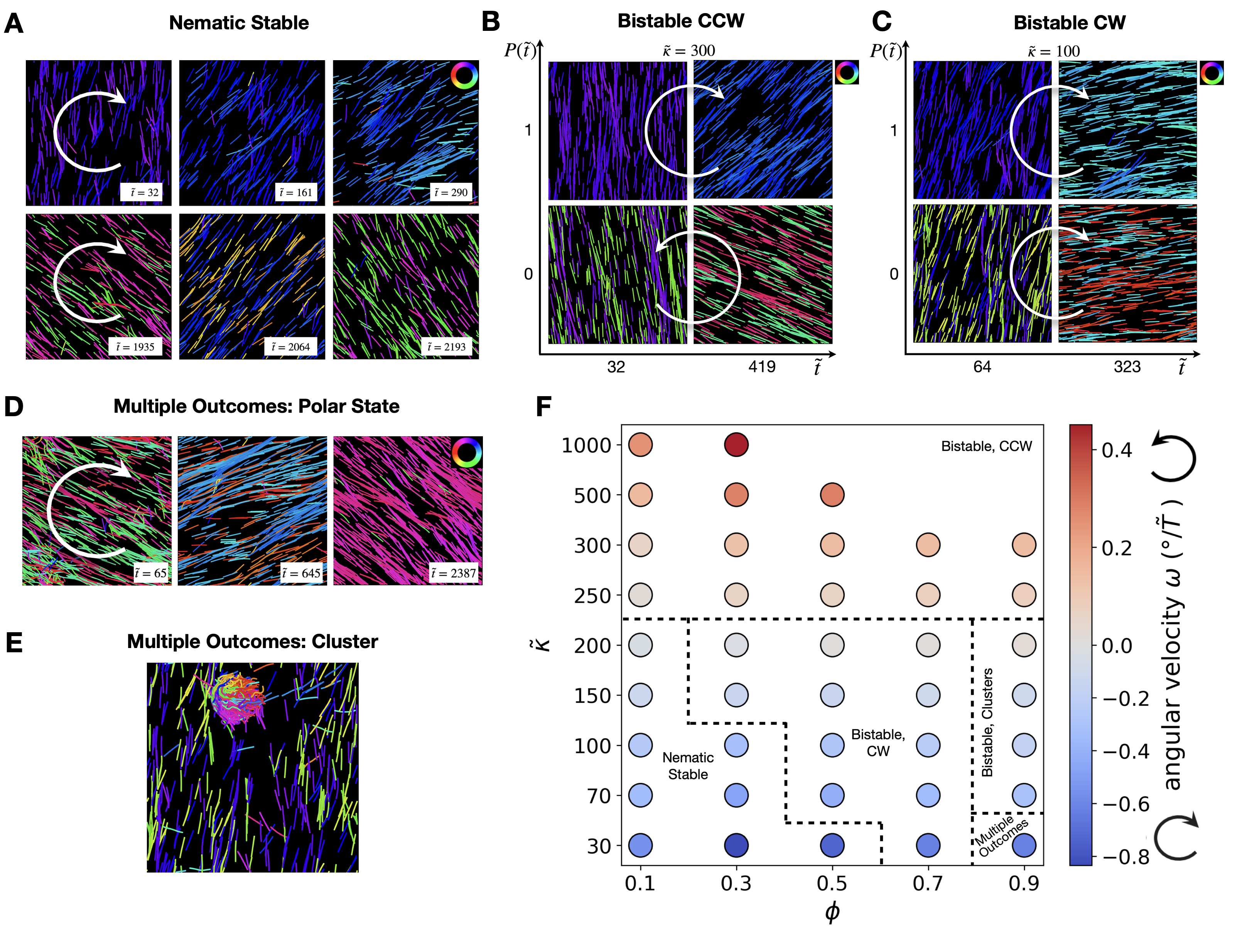}
    \caption{\textbf{A.} Snapshots of time evolution in a typical simulation from a polar initialization toward a nematic steady state for $\alpha = 0.1$rad CCW, $\tilde \kappa = 30$, $\phi= 0.3$, $L_x = 150$.  \textbf{B},\textbf{C}. Snapshots of the two different types of bistable steady-state scenarios. \textbf{B.} For higher $\tilde{\kappa}$ values, a system initialized as a LRO nematic (bottom row) exhibits CCW collective rotation, whereas a system initialized as a LRO polar (top row) rotates CW. \textbf{C.} For lower $\tilde{\kappa}$ values, CW collective rotation is observed for both nematic (bottom row) and polar (top row) initial conditions. Parameters are $\phi = 0.5$, $L_x = 150$ and $\alpha = 0.1$rad CCW. \textbf{D.} Snapshots of a simulation whose director rotates CW with LRO polar order as its steady state; $\tilde{\kappa} = 30$, $\phi = 0.9$. \textbf{E.} Example of a motile cluster formed within a previously uniform LRO nematic. Motile clusters can either be transient or can grow to eventually consume all filaments. \textbf{F.} Phase diagram showing the symmetries of observed steady states and their angular velocities $\omega$ (units of $^{\circ}/\tilde{T}$ where $\tilde{T} = l/v_0$), as a function of $\phi$ and $\tilde{\kappa}$. The five stability regimes described in this figure are delineated by dashed lines.}
    \label{fig:5}
\end{figure*}

As noted above, 
the time cholesteric state's stabilization of nematic order stands in contrast with achiral systems in the same model, where nematic order generically decays into a polar steady state \cite{athani2023symmetry}. This raises the question of whether chiral self-propulsion makes polar-ordered states absolutely unstable, or whether they merely make nematic states stable. Furthermore, if polar-ordered states are long-lasting, do they rotate coherently like their nematic counterparts? To gain insights into these questions, 
we initialized systems with perfect polar order and analyzed the resulting dynamics. We held the active force skew angle $\alpha$ fixed at $0.1$ rad, and varied the bending rigidity $\tilde \kappa$ and area fraction $\phi$. With $\tilde \kappa = 30$ and $\phi = 0.3$, we found that initial polar order decays steadily into a nematically-ordered steady state Fig.~\ref{fig:5}\textbf{A}. The nematic order parameter remains high throughout the process, as expected for either a nematic or polar state, while the polar order parameter decreases over time to approximately zero (SI Fig.~S13\textbf{A},\textbf{B}). From this, we conclude that the collective rotation of the time cholesteric can destabilize long-range polar order and stabilize long-range nematic order, reversing the decay of nematic to polar alignment seen in the achiral system. The angular speed of the globally averaged orientation increases in magnitude and then saturates as polar order decays (SI Fig.~S13\textbf{B} inset).

To better understand this change in stability introduced by nonzero $\alpha$, 
we next examined systems initialized with perfect nematic order and zero polar order, with $\alpha$ incrementally reduced from $0.1$ to $0.01$ across many simulations. We find that the behavior of the achiral system in  Ref.~\cite{athani2023symmetry} is gradually recovered (SI Fig.~S13\textbf{E}): as $\alpha$ decreases, polar order regains stability over nematic order, first at small area fractions $\phi$ and then at all $\phi$. For example, with $\alpha = 0.03$ rad, $\tilde \kappa = 30$, and $\phi=0.3$, we find that a system initialized in a nematic state eventually reaches LRO polar order (SI Fig.~S13\textbf{C},\textbf{D}). Just as decaying polar order was accompanied by an increasing angular speed with $\alpha = 0.1$, here the angular speed decreases as polar order rises over time for $\alpha =0.03$ (SI Fig.~S13\textbf{D}). 

The area fraction of filaments influences the steady state at intermediate $\alpha$: With the active skew angle held fixed at $\alpha = 0.03$ rad, increasing $\phi$ from $0.3$ to $0.5$ changes the steady state from LRO polar to LRO nematic (SI Fig.~S13\textbf{E}). The relative stability of nematic and polar order thus depends strongly on both bending rigidity $\tilde \kappa$ and area fraction $\phi$, even at fixed $\alpha$. In particular, with $\alpha = 0.1$ rad, the introduction of chiral activity destabilizes the LRO polar state in favor of the LRO nematic only for sufficiently flexible filaments $\tilde \kappa \lesssim 200$ and sufficiently low $\phi$.

Our findings are summarized in a phase diagram, Fig.~\ref{fig:5}\textbf{F}, that represents the main active morphologies for various values of $\tilde \kappa$ and $\phi$, at a fixed active skew angle of $\alpha = 0.1$ rad. We find that this parameter space is divided into five qualitatively distinct steady-state possibilities. First, in the \textit{Nematic stable} region (low $\phi$ and low $\tilde \kappa$), the steady state is the LRO nematic state regardless of the initial condition. Here, the steady-state angular velocity $\omega$ is comparable between systems initialized with isotropic or strong polar orders. Second, in the \textit{Bistable region CCW} ($\tilde \kappa \gtrsim 200$), the steady state depends on the initial condition. If the initial condition is LRO polar, then the system remains strongly polar and rotates CW. However, when the system is initialized in the isotropic state \textit{or} as a perfect nematic, the steady state is LRO nematic, with a director that rotates CCW. Thus, remarkably, there are parameters for which a single system can exhibit oppositely rotating steady states depending on the initial condition (Fig.~\ref{fig:5}\textbf{B}). Third, there is also a \textit{Bistable region CW} where, for intermediate values of $\phi$ and with $\tilde \kappa \lesssim 200$, we again observe the bistability of LRO nematic and LRO polar states, depending on initial conditions. Here, however,  both the LRO nematic and the LRO polar states exhibit CW collective rotation (Fig.~\ref{fig:5}\textbf{C}). Fourth, \textit{Bistable with motile clusters} is observed for high area fractions of $\phi = 0.9$ and $ 30 <\tilde \kappa < 200$. Here, bistable CW rotating LRO states can lead to the formation of slow-moving clusters, which we interpret as a form of motility-induced phase separation \cite{cates2015motility} caused by the crowding of counter-moving filaments at this high area fraction. Fifth, and finally, the \textit{Multiple outcomes region} accounts for high occupancy by especially flexible filaments ($\phi = 0.9$ and $\tilde \kappa = 30$), where the outcome is not precisely defined by the conditions and the system can converge to various outcomes such as a polar state (Fig.~\ref{fig:5}\textbf{D}, Fig.~S14, and SI Movie S7), or clusters (Fig.~\ref{fig:5}\textbf{E}).

Interestingly, wherever we observed a dynamically stable LRO polar state, its rotation was CW, opposite to the handedness of the chiral self-propulsion. This is true even in the regime we label \textit{Bistable, CCW}, whose name refers to the sense of rotation of the LRO nematic state. Because collisions in an LRO polar state are expected to occur mostly between nearly parallel filaments, aligning events dominate over crossovers to a greater degree than in the LRO nematic state. The association of polar order with exclusively CW rotation in the collective rotation thus indirectly corroborates our argument in Sec.~\tunableSectionName that aligning events promote rotation that is opposite in handedness to the chiral bias of the active force. 

Much of this parameter space extends beyond our current experimental range, but these findings point to rich behaviors that may be experimentally probed in future active matter investigations.

\section*{Discussion and conclusions}
Left-right symmetry breaking is an important step during morphogenesis. Previous works have hypothesized that the inherent chirality of cytoskeletal filaments such as microtubules and actin  may play a role in this symmetry breaking \cite{Houliston1994,Danilchik2006,Lobikin2012,Tran2012,Henley2012,Tee2015,McDowell2016}. The coherently rotating active nematic state, or ``time cholesteric,'' of microtubule-kinesin gliding assays offers a simple testbed for theories of how chiral structure at the molecular scale is transmitted to mirror symmetry breaking in larger-scale dynamics. 

Our experimental results qualitatively confirm earlier observations of CCW rotation of the nematic director in collective microtubule gliding assays. Quantitatively, previously reported rotation rates span approximately one order of magnitude, from \(0.008^\circ/\mathrm{s}\) \cite{Tanida2020} to \(0.02^\circ\)–\(0.07^\circ/\mathrm{s}\)  \cite{kim2018}. The rotation rates measured on GMPCPP-microtubules in our study fall within this range, supporting the consistency of the phenomenon across different experimental conditions. While some earlier studies attributed the rotation behavior to a preferred signed curvature of isolated  microtubule gliding trajectories, we have presented evidence for microtubule collisions as the main driver of the collective rotation. This evidence came from our single-filament assays, together with the observed anti-correlation between collective rotation rate and nematic coherency.

Our Brownian dynamics simulations provide significant insights into time cholesteric behavior. Motivated by protofilament supertwist  in microtubules with protofilament numbers $N \neq 13$ \cite{Hyman1995} and the fact that kinesin-1 exerts forces along the direction of the microtubule protofilament \cite{Ray1993, Nitzsche2008}, we modeled chiral activity as self-propulsion directed at a fixed, positive skew angle $\alpha$ to the local tangent. Our simulations demonstrated that chirality in active self-propulsion is sufficient to produce the collective behavior of time cholesterics in the bulk. This indicates  that chirality in the filaments' average conformation is not a necessary component of time cholesterics. Our model is versatile and includes a regime of low bending rigidity in which individual filaments rotate with a fixed handedness, but this does not occur when $\tilde \kappa$ has the order of magnitude expected to apply to microtubules in gliding assays.  Furthermore, our analysis of these simulation results supports the picture of collective rotations arising from collisions between filaments. The importance of filament collisions, as opposed to individual filament turning, to the time-cholesteric rotation is indirectly corroborated by the finding in Ref.~\cite{kim2018} that the angular velocity changes in response to changes in the concentration of depletants, which strongly modulate the filament-filament interactions \cite{Huber2018, Tanida2020}. Such collision behavior has previously been characterized extensively for lipid-based gliding assays \cite{grover2016transport}. 


Our model predicts that the sense of rotation in the time cholesteric state can be reversed in two ways. The first is obvious: changing the handedness of the chiral self-propulsion by taking the active skew angle $\alpha$ to $-\alpha$ will result in a mirror-inversion of the dynamics. The second is a highly non-trivial emergent feature of our model, in which the time cholesteric rotates with the same sense as $\alpha$ when $\tilde \kappa$ is greater than a certain threshold value $\tilde \kappa_c$, but with the opposite sense below it. We found evidence for the latter reversal in our gliding assay experiments on lipid substrates, in which CCW collective rotation was observed both in GMPCPP microtubules ($\alpha>0$, higher bending rigidity) and in taxol-stabilized microtubules ($\alpha<0$, lower bending rigidity). In the taxol case, the collective rotation with the sense opposite to the chiral self-propulsion is difficult to understand without a model that predicts a second sign reversal of angular velocity due to other parameters, in our case the bending rigidity.

While changing the protofilament number $N$ can change the handedness of the supertwist, it does not change the left-handed nature of the lateral bonds between tubulin dimers on neighboring protofilaments. Thus, kinesin motor sidestepping, which is biased leftward \cite{mitra2018directionally}, may contribute to making the effective $\alpha$ positive for all $N$, as argued by some previous works  \cite{kim2018, Meiner2024}. It is therefore plausible that this helicity, which is the same for taxol-stabilized and GMPCPP microtubules, determines the sense of the collective rotation. Indeed, all previous experimental studies of nematic order in microtubule-kinesin gliding assays report a CCW rotation, irrespective of the biochemical details. We cannot rule out this hypothesis. If true, this would invalidate our proposal that taxol-stabilized microtubules rotate opposite to the handedness of their chiral self-propulsion. 

However, our  model remains applicable in this scenario, so long as we are free to re-classify taxol-stabilized microtubules as belonging to the high-$\tilde \kappa$ regime that rotates with the same sense as the handedness of the self-propulsion. Furthermore, our model of an active force directed at a small angle $\alpha$ away from the tangent remains a reasonable hypothesis as the time-averaged kinesin motion is primarily along the tube axis and only slightly biased by occasional side-stepping.

In our experiments, we investigated how collective behavior in microtubule gliding assays relates to motor attachment at the substrate interface. On glass substrates, we observed that the emergence of collective nematic order strongly depends on surface motor density. This finding contrasts with previous studies, which suggested motor density has a minimal impact on the outcome of microtubule collisions \cite{Sumino2012,Inoue2015}. However, when comparing the actual surface densities used in those studies with ours, we find that our motor densities are lower by more than an order of magnitude, potentially accounting for the discrepancy. Although our model does not explicitly incorporate motor density, we expect that increasing the motor density corresponds to increasing the active force in our model, which should have qualitatively the same result as decreasing the bending rigidity.



It would be valuable to test our model using experiments with only canonical 13-protofilament microtubules. However, this approach would necessitate employing unstabilized microtubules to maintain a narrow protofilament-number distribution, which is experimentally challenging due to difficulties in preserving stable microtubule lengths throughout the extended experimental time-frames required for this analysis. Consequently, this approach was deemed impractical within the present study. Previous work using gliding assays of non-stabilized microtubules grown from stabilized seeds reported no rotation of the nematic order \cite{kuvcera2022actin}. 
Another possible direction for experimentally testing our predictions is the generation of initial microtubule arrangements with polar order; however, achieving such polar ordering is currently an unresolved experimental challenge in gliding assays.


Our simulations further predict that chiral activity significantly affects the interplay of polar and nematic symmetries in the LRO steady state. In a previous computational work~\cite{athani2023symmetry} studying collective motions of active, achiral, semiflexible filaments, the LRO nematic was shown to be a transient state whereas the LRO polar state was the steady state. In contrast, with chiral activity, we have shown that at lower bending rigidities $\tilde \kappa$ and area fractions $\phi$, the LRO nematic is the steady state (Fig.~\ref{fig:5}\textbf{A}). This presents an interesting contrast with the findings in Refs.~\cite{Afroze2021,Hiraiwa2022} that chirality favors polar order over nematic order, a distinction likely owing to the easy mutual penetrability of filaments in our model. On the other hand, rotation was found to stabilize long-range orientational order, both nematic and polar, in a hydrodynamic model of rotating active liquid crystals \cite{maitra2019spontaneous}.
 It is interesting to compare this predicted hydrodynamic stability of both symmetries with the remarkable dynamical bistability exhibited by our numerical model, as we increase  $\tilde \kappa > 200$ or $\phi$; the steady state has long-range polar order if the system is initialized with that symmetry, and long-range nematic order if it is initialized with nematically ordered or isotropically disordered filaments (Fig.~\ref{fig:5}\textbf{B},\textbf{C}).  

At very high area fractions, the uniform-density LRO states give way to density inhomogeneities in the form of motile clusters. Sometimes a LRO polar steady state can also be seen (Fig.~\ref{fig:5}\textbf{D}) at high $\phi$. All these states exhibit the same change in sense of rotation--from opposite to like handedness--relative  to the chiral self-propulsion as $\tilde \kappa$ increases through a threshold value $\tilde \kappa_c$ where the rotation vanishes. 

Our results demonstrate the propagation of molecular-scale chiral properties into system-scale chiral dynamics, a phenomenon strikingly reminiscent of cortical microtubule reorganization observed in plant cells \cite{wasteneys2002microtubule, chan2007cortical}. Existing studies in model plants indicate that this reorganization primarily relies on microtubule assembly dynamics and the activity of microtubule-severing protein complexes \cite{lindeboom2013mechanism}, whereas molecular motors have been previously reported as not playing a significant role \cite{vos2004microtubules}. However, given the limited evidence from other model systems, it remains an open question whether motor proteins might contribute more generally to cortical microtubule transport and reorganization. Should such involvement prove common, our findings could provide a feasible and mechanistically clear pathway to explain motor-driven reorientation of cortical microtubule arrays in plants. Beyond plant cells, we speculate that our findings could be relevant to the organization of microtubules in other cellular systems, potentially influencing morphogenesis. Finally, we hope that the variety of dynamically bistable states identified by our model will help to broadly inspire further explorations of chiral mechanisms for switchable dynamics in both biological and synthetic active matter.

\matmethods{

\paragraph{Protein preparation:} Bovine brain tubulin was isolated by four cycles of polymerization and depolymerization \cite{SHELANSKI1973}. The protein was further purified from any associated proteins by utilizing cation exchange chromatography \cite{VANTARD1994}. A part of the purified tubulin was fluorescence-labeled by Atto488 fluorophore or rhodamine through the NHS ester coupling. A truncated kinesin-1-GFP motor with a 6x histidine tag was expressed in E. coli and purified as previously described \cite{Case1997}.

\paragraph{Microtubule polymerization:} Microtubules were polymerized from 4 mg/ml tubulin in either PEM100 (100mM PIPES, 1mM EGTA, 1mM MgSO$_4$, pH6.9) with 1mM GTP or in BRB80 (80mM PIPES, 1mM EGTA, 1mM MgCl$_2$, pH6.9) with 1mM GMPCPP and 1mM MgCl$_2$. For taxol-stabilized microtubules, the GTP-containing solution was incubated at $37^\circ$C for 20 minutes, followed by addition of 50\textmu M taxol and a further 20-minute incubation. Samples were then centrifuged at 14,000×\emph{g} for 10 minutes at $25^\circ$C, and pellets were resuspended in the original volume of PEM100 with 50\textmu M taxol. For GMPCPP microtubules, the mixture of labeled and unlabeled tubulin with a final labeled fraction of tubulin around 10\% was incubated at $37^\circ$C for 2 hours and centrifuged at $18,000\times g$ for 30 minutes. The resulting pellet was resuspended in BRB80 supplemented with 10\textmu M taxol. For experiments on lipid bilayers, taxol-stabilized microtubules polymerized from 30\% rhodamine-labeled tubulin (with 70\% unlabeled) and from 100\% unlabeled tubulin were mixed to obtain a final labeling fraction of 6.9–13\%. Only the labeled microtubules were visible by fluorescence microscopy. Preparations were stored at room temperature and used within 24 hours.

\paragraph{Experiments on glass:} The glass coverslips were subjected to sonication at room temperature in 1M NaOH for 1 hour. Then, the coverslips were washed with Milli-Q water and placed in a beaker filled with ethanol and slightly agitated for 30 minutes. Finally, the coverslips were thoroughly rinsed with Milli-Q water and dried just before use. The channel for imaging was created by connecting these NaOH-activated glass coverslips with double-sided tape that had a channel cut into it. For microtubule gliding on immobilized kinesin motors, the surface of the channel was first functionalized with GFP antibodies (Invitrogen, A-11122, 100 $\mu$g/ml in 10 mM Hepes pH 7.2, 5 mM MgCl2, 1 mM EGTA, and 50 mM KCl) that were incubated in the channel for 3 minutes. The channel was then passivated by BSA (Bovine serum albumin, 1\% w/v) for 5 minutes. Kinesin-1-GFP molecules (6 or 60 $\mu$g/ml in wash buffer: 10 mM HEPES buffer (pH 7.2), 16 mM PIPES buffer (pH 6.8), 50 mM KCl, 5 mM MgCl2, 1 mM EGTA, 20 mM dithiothreitol (DTT), 3 mg/ml glucose, 20 $\mu$g/ml catalase, 100 $\mu$g/ml glucose oxidase, and 0.3\% w/v BSA) were introduced into the channel by perfusion of 2 channel volumes and left to attach specifically to the antibodies for 3 minutes. After the channel was perfused with three volumes of wash buffer, microtubules at dilutions ranging from 1:10 to 1:300 in the wash buffer were introduced into the channel and allowed to bind with kinesin motors for one to five minutes. After the channel was perfused with three volumes of wash buffer again, the imaging ATP-containing buffer was introduced (10 mM HEPES pH 7.2, 16 mM PIPES pH 6.8, 50 mM KCl, 5 mM MgCl2, 1 mM EGTA, 20 mM dithiothreitol (DTT), 3 mg/ml glucose, 20 $\mu$g/ml catalase, 100 $\mu$g/ml glucose oxidase, 1 mM ATP, 1 mM GTP, 0.3\% w/v BSA, and 0.327\% w/v methylcellulose (63 kDa, Sigma-Aldrich, M0387)). To avoid any evaporation during the imaging process, the channel was sealed with a capillary tube sealant (Vitrex). The surface density of kinesins was evaluated by comparing their fluorescence signal in the channel with the intensity of individual dimeric kinesin-1-GFP motors attached to a neighbouring channel at much higher dilution \cite{Kucera2022Actin}.

\paragraph{Lipids preparation:} For experiments on lipid membranes, we used a mixture of 97\% w/v L-$\alpha$-phosphatidylcholine (10mg/ml, EggPC, Avanti, 840051C), 2\% w/v 1,2-di-(9Z-octadecenoyl)-sn-glycero-3-[(N-(5- amino-1-carboxypentyl)iminodiacetic acid)succinyl] (nickel salt) (10mg/ml, 18:1 DGS-NTA(Ni), Avanti, 790404C) and 1\% w/v 1,2-dioleoyl-sn-glycero-3-phosphoethanolamine-Atto 390 (1mg/ml, DOPE-Atto 390, AttoTec, AD 390-161) that were mixed in glass tube, dried with nitrogen gas, incubated overnight, and finally hydrated in the SUV buffer (10 mM Tris (pH 7.4), 150 mM NaCl, 2 mM CaCl2). This mixture was sonicated for 10 mins on ice, then centrifuged for 10 mins at $20,238 \times g$, and the supernatant was stored at 4$^\circ$C until use.

\paragraph{Experiments on lipids:} The glass coverslips were rinsed extensively with Milli-Q water and ethanol and then sonicated for 30 minutes in 2\% Helmanex III solution at 60$^\circ$C. Then, the coverslips were rinsed with Milli-Q water extensively and dried right before use. A microfluidic channel was created by attaching two coverslips together with double-sided tape that was pre-cut to the channel shape. Once assembled, the channel was perfused with a solution containing small unilamellar vesicles made of a lipid mixture (as described above). The lipid mixture was diluted to approximately 0.5 mg/ml (standard concentration for supported lipid bilayer) or 0.1 mg/ml (diluted concentration for fragmented lipid bilayer) in an SUV buffer. The vesicles spread over the glass surface, forming a supported lipid bilayer during the 10-minute incubation period. The channel was then washed by four channel volumes of the SUV buffer, followed by 1\% w/v bovine serum albumin (BSA) in HKEM buffer (10 mM HEPES-Na (pH 7.2), 50 mM KCl, 1 mM EGTA and 5 mM MgCl2), that was kept in the channel for 5 minutes to passivate any remaining exposed glass surface. Kinesin-1-GFP molecules (60 $\mu$g/mL in wash buffer) were introduced into the channel by perfusion of 2 channel volumes. They were left in the channel for 5 minutes to attach specifically through the His-tag to nickel contained in phospholipids. The channel was then perfused with three volumes of wash buffer, followed by microtubules (1:10 in the wash buffer) that were allowed to bind with kinesin motors for five minutes. Next, the channel was perfused with three volumes of wash buffer again. Finally, the imaging ATP-containing buffer was introduced, and the channel was sealed with Vitrex.

\paragraph{Imaging and image analysis:} Microtubules, kinesin motors, and phospholipids were imaged using Total Internal Reflection Fluorescence (TIRF) microscopy. The inverted microscope used was Eclipse Ti (Nikon), equipped with a 100x 1.49 N.A. oil immersion objective (UApo N, Olympus), 491 nm and 405 nm lasers (Optical Insights) and an iLas2 dual laser illuminator (Roper Scientific) for exciting the GFP, Atto488, and Atto390 fluorophores. Diffusivity of the lipid bilayer and the kinesin motors within was verified before each experiment by observing the fluorescence recovery after photobleaching (FRAP) of a circular patch (diameter 11.7 $\mu$m) of the bilayer by an ultraviolet (UV) laser. The Evolve 512 EMCCD camera (Photometrics) was used to capture the images. The imaging and the FRAP sequence were controlled by the Metamorph software (v. 7.7.5, Universal Imaging). In addition, conventional fluorescence microscopy was employed for selected experiments using a DM 2500P upright microscope (Leica Microsystems Inc.) equipped with 20x or 40x objectives.Fluorescence movies were recorded under low-light conditions using an ORCA-Flash4.0 LT+ digital CMOS camera (Hamamatsu). Raw images were processed with an automatic brightness and contrast adjustment routine, as implemented in Fiji \cite{schindelin2015imagej}, to enhance clarity for final presentation. Orientation analysis was performed using OrientationJ \cite{puspoki2016}. Tracking was performed using FIESTA \cite{ruhnow2011tracking} and custom-made scripts (see Supplementary Information). 

Extraction of Multiple Fields of View from Raw Images: For taxol stabilized microtubules on lipid membrane Fig \ref{fig:4}\textbf{D}, time-lapse fluorescence images of sizes (1392 x 1040 pixels, 1$\mu$m = 3.1px)  were cropped to exclude edge regions (to avoid any boundary effects). A margin of about 2 times the mean microtubule (MT) length ( 43px, approximately 2 times 7$\mu$m) was excluded from each side. From the interior region (1349 x 997 pixels), we extracted 15 non-overlapping square fields of view of size 254 x 254 pixels (81.9$\mu$m), retaining all frames across time.
 
Measurement of nematic order parameter (coherency): The nematic order parameter was time-averaged per file, then ensemble-averaged across n experiments. Error is reported as the standard error of the mean (SEM).
} 

\showmatmethods{} 

\acknow{The authors thank Poorvi Athani for assistance in constructing 3D graphics, Magali Orhant-Prioux (CEA) for technical assistance with lipid bilayers, Clothilde Utzschneider (CEA) for help with initial experimental work, and Stefan Diez (TU Dresden) for the fruitful discussion of these results. 

This work was supported by the European Research Council, Advanced Grant 741773 (AAA) to LB. OK was supported by European Research Council, Consolidator Grant 771599 (ICEBERG) to Manuel Th\'ery (CEA). OK was also partially supported by P\^{o}le emploi (7820342X) and the Czech Science Foundation, grant 25-17332S. The imaging platform at CEA is supported by the Laboratory of Excellence Grenoble Alliance for Integrated Structural and Cell Biology (LabEX GRAL)(ANR-10-LABX-49-01) and the University Grenoble Alpes graduate school (Ecoles Universitaires de Recherche, CBH- EUR-GS, ANR-17-EURE-0003). This material is based upon work supported in part by the National Science Foundation under Grant No.~DMR-2225543. This work was also supported by the National Science Foundation NSF-CREST: Center for Cellular and Bio-molecular Machines at UC Merced (NSF-HRD-1547848 and EES-2112675 to AG). AGK, NS and AG also acknowledge partial support from the NSF Center for Engineering Mechanobiology grant CMMI-154857 and computing time on the Multi-Environment Computer for Exploration and Discovery (MERCED) cluster at UC Merced (NSF-ACI-1429783). 
}

\bibliography{chiral.bib}%

\supportinginformation

\section{Effect of increased motor density in  gliding assay on a glass substrate}

Doubling the kinesin motor density to approximately 34.8 motors per square micrometer led to inconsistent formation of collective nematic order of microtubules with density of $0.12 \pm 0.01$ filaments per square micrometer ($n = 4$ experiments, Fig.~S2). In cases with nematic order ($n = 2$ experiments out of 4), the rotation of the nematic director was $0.0017 \pm 0.0002$ degrees per second, even though individual microtubules under these conditions have a turning rate of $0.055$ (95\% confidence interval bounded by $0.03611$ and $0.07293$) degrees per second ($p = 0.014$, Mann–Kendall test of the significance of the trend, $n = 300$ tracks from 14 experiments). 

\begin{figure*}[h!]
    \centering
    \includegraphics[width=17cm]{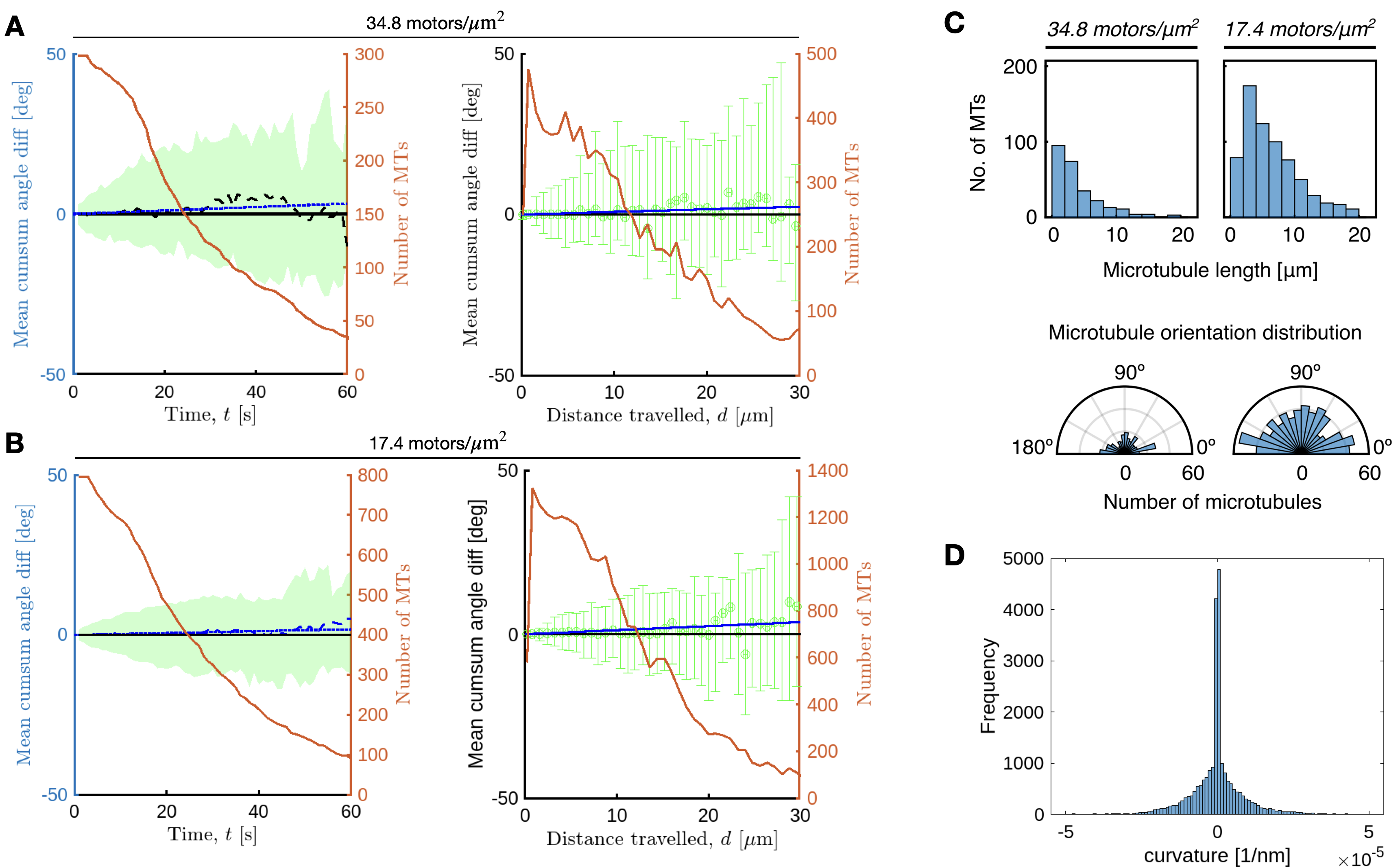}
    \caption{Gliding of individual GMPCPP-stabilized microtubules on a glass substrate. \textbf{A} and \textbf{B.} Cumulative angle difference of gliding microtubule as a function of time (left) and
distance travelled (right). The green curve represents the median, and the blue line is a linear fit weighted by the
number of contributing microtubules. The shading and whiskers’ extent represent the lower and upper quartiles.
The red curve shows the number of microtubules contributing to the statistics. \textbf{C.} Distribution of microtubule lengths and their
initial orientation. \textbf{D.} Curvature distribution of single GMPCPP-stabilized microtubules in a dilute gliding assay.}
    \label{SI-single_MT_glass}
\end{figure*}

\begin{figure*}[h!]
    \centering
    \includegraphics[width=17cm]{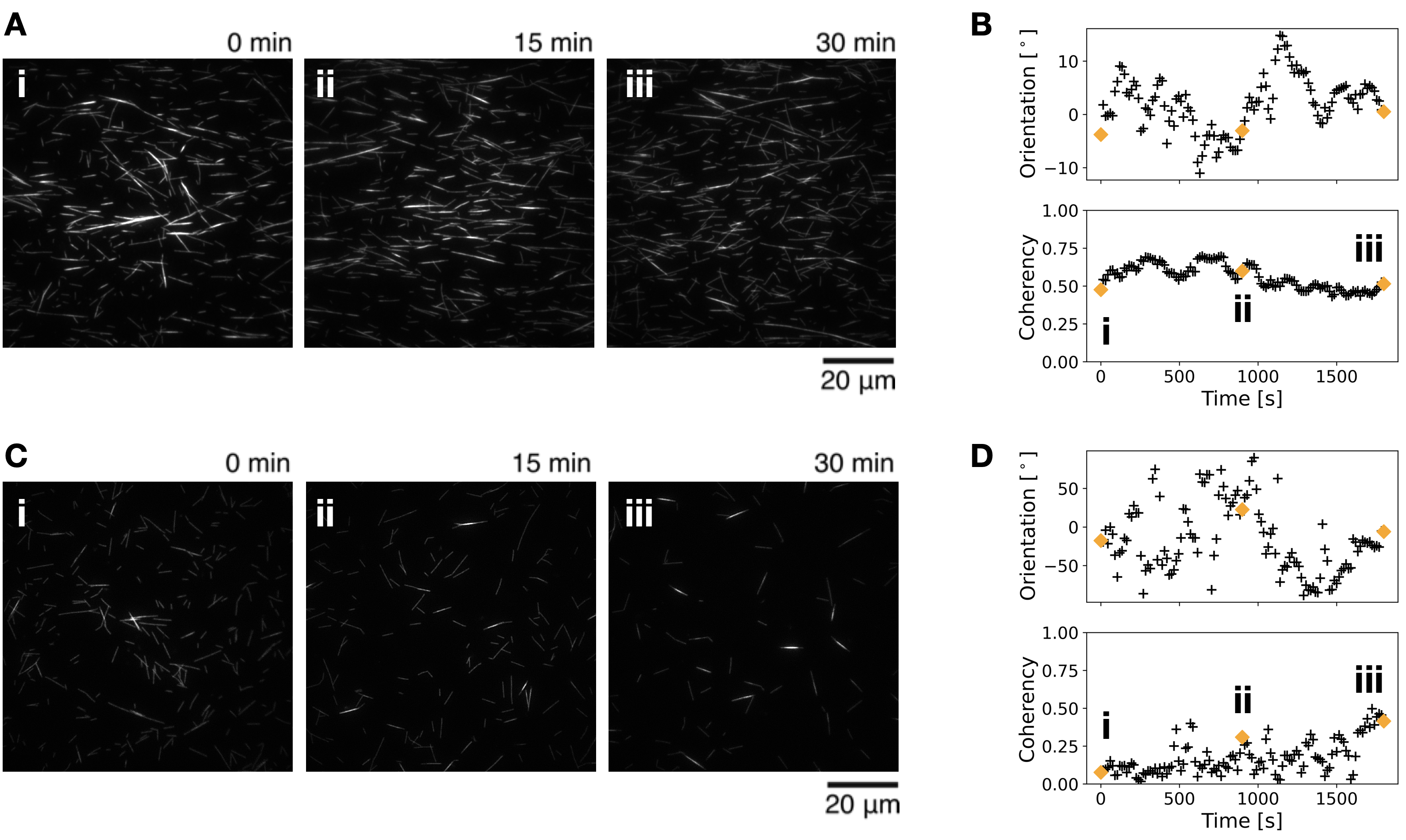}
    \caption{Inconsistency in the formation of a LRO nematic state of GMPCPP-stabilized MTs on a glass substrate with high kinesin-1 motor density of 34.8 motors/$\mu$m$^2$. \textbf{A.} Snapshots from the experiment that forms a LRO nematic but does not rotate. \textbf{B.} Orientation and coherency of the experiment shown in \textbf{A}. \textbf{C.} Snapshots from an experiment under the same conditions of \textbf{A}, but no LRO nematic state formed. \textbf{B.} Orientation and coherency of the experiment shown in \textbf{C}.}
    \label{SI-glass_high_motor}
\end{figure*}

\section{Binary collision simulations}

\begin{figure*}[h!]
    \centering
    \includegraphics[width=17cm]{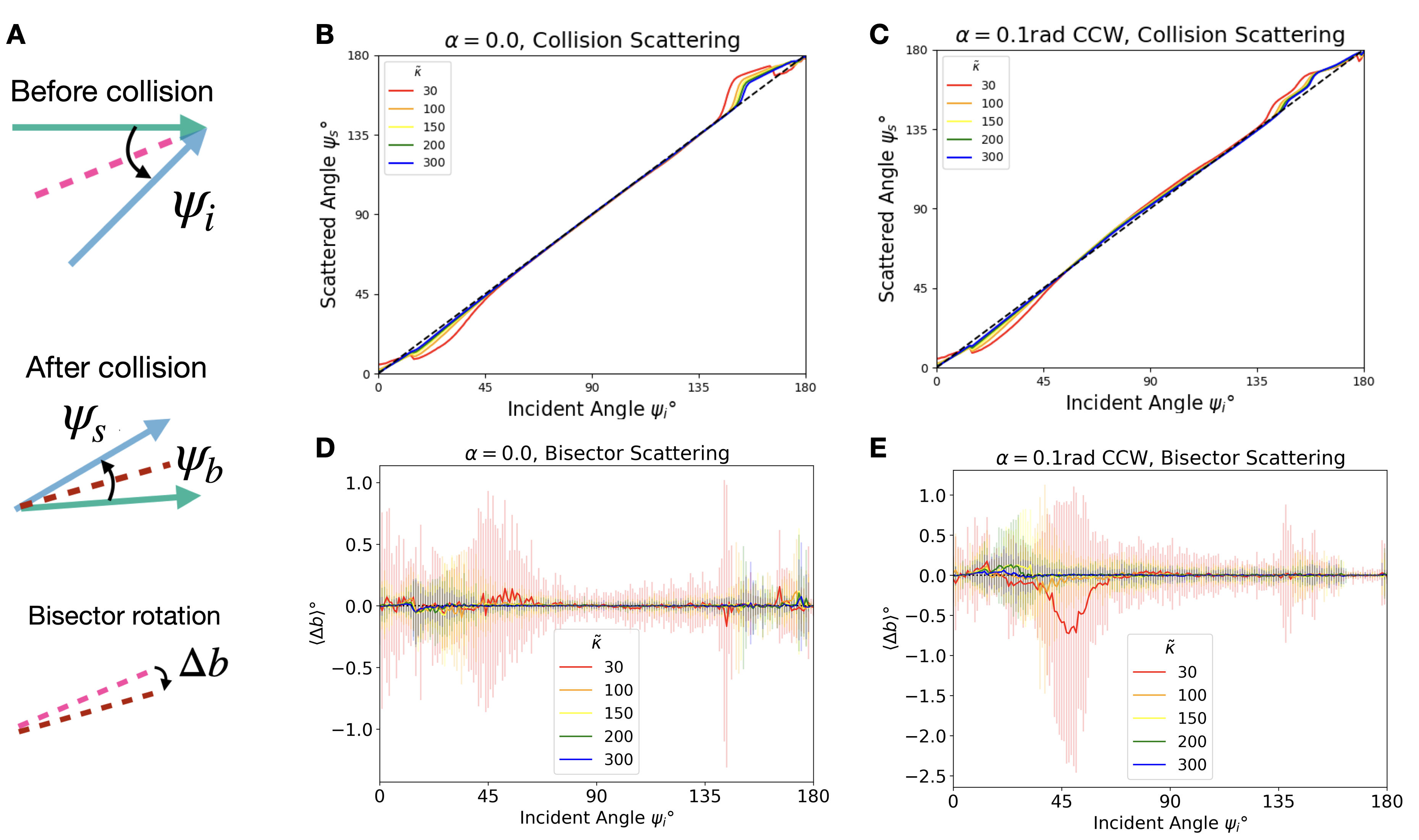}
    \caption{\textbf{A.} Illustration of a pair collision event and the associated angles measured in the simulation. \textbf{B,} \textbf{C.} Average scattered angle as a function of the incident angle $\psi_i$ for (C) $\alpha = 0.0$ and (D) $\alpha = 0.1$rad. \textbf{D.} Average rotation of the filament pair bisector for each incident collision angle demonstrating no net rotation when chiral offset $\alpha = 0.0$. \textbf{E.} A net CW rotation is seen for low bending rigidity of $\tilde \kappa = 30$ which is opposite in handedness to the chiral offset $\alpha = 0.1$rad CCW on each filament, while for higher $\tilde \kappa$ there is a slight excess of $\Delta b$ with the same handedness as $\alpha$. Error bars show the standard deviation.}
    \label{SI-binaryMT}
\end{figure*}

We conduct a systematic analysis of collisions of a pair of simulated filaments with chiral activity to investigate how these interactions may lead to the collective rotation of a LRO nematic state. Collisions are simulated for a range of incident angles $\psi_i \in [0,2\pi]$. Each collision is sampled $n=200$ times for $n_L=16$ equidistantly spaced head-to-body collisions along the length of each filament. After the collision ends, we record the final in-plane angle $\psi_s$ of each filament. A collision is said to have ended when all pairs of beads on the two filaments are farther apart than the range of the  WCA interaction potential. Both $\psi_i$ and $\psi_s$ are relative angles between the two filaments, and do not refer to a choice of coordinate system. From this data, we calculate both the change in the angle difference between the filaments' orientations, $\psi_s - \psi_i$, and their net rotation, as measured by the change $\Delta b$ in the bisector $\psi_b$ between their individual orientations; see Fig.~\ref{SI-binaryMT}\textbf{A} for schematic illustrations. 

We plot the measured scattered angle $\psi_s$ as a function of incident angle $\psi_i$ for a range of bending rigidities, both for the $\alpha=0$ achiral model (Fig.~\ref{SI-binaryMT}\textbf{B}) and for $\alpha =0.1\,\mathrm{rad}$ CCW (Fig.~\ref{SI-binaryMT}\textbf{C}). For both $\alpha$ values and at all $\tilde \kappa$ values, when the incident angle $\psi_i$  is $\lesssim 45^\circ$, we observe a contraction in the angle between the filaments ($\psi_s < \psi_i$). This indicates that the collision makes the filaments more parallel, as required by either polar or nematic ordering. Conversely, for $\psi_i$ near $180^\circ$, we observe an expansion in the angle between the filaments ($\psi_s > \psi_i$). This indicates that the collision brings the filaments into more nearly antiparallel alignment, consistent with a trend toward nematic but not polar order. The deviation of $\psi_s$ from $\psi_i$ generally decreases as $\tilde \kappa$ increases, consistent with the finding in Ref.~\cite{athani2023symmetry} that polar order arises more slowly for larger $\tilde \kappa$. 

The contribution of each collision to the global rotation is approximated by the rotation $\Delta b$ of the bisector between the filaments. As a check, we confirm in Fig.~\ref{SI-binaryMT}\textbf{D} that the achiral filaments have an approximately zero mean $\Delta b$ (that is, deviations from zero show no structure and are small compared to the standard deviation) for all bending rigidities and incident angles. In contrast, Fig.~\ref{SI-binaryMT}\textbf{E} shows that the deviation of $\Delta b$ from zero as a function of $\psi_i$ has significant structure, though in different ways for different bending rigidities. For $\tilde \kappa = 30$, the most prominent feature is a strongly negative $\Delta b$ range near $\psi_i = 45^\circ$. For larger values of $\tilde \kappa$, the deviation from zero is concentrated in a range of positive values centered around $\psi_i \approx 30^\circ$. This association of smaller $\tilde \kappa$ with CW rotation and larger $\tilde \kappa$ with CCW rotation is qualitatively consistent with our findings in the bulk for the $\tilde \kappa$-dependence of angular velocity $\omega$ (main text Fig.~3), suggesting that the change in handedness of the bulk rotation arises from a change in the average handedness of pair collisions.

\begin{figure}[t]
    \centering
    \includegraphics[width=17cm]{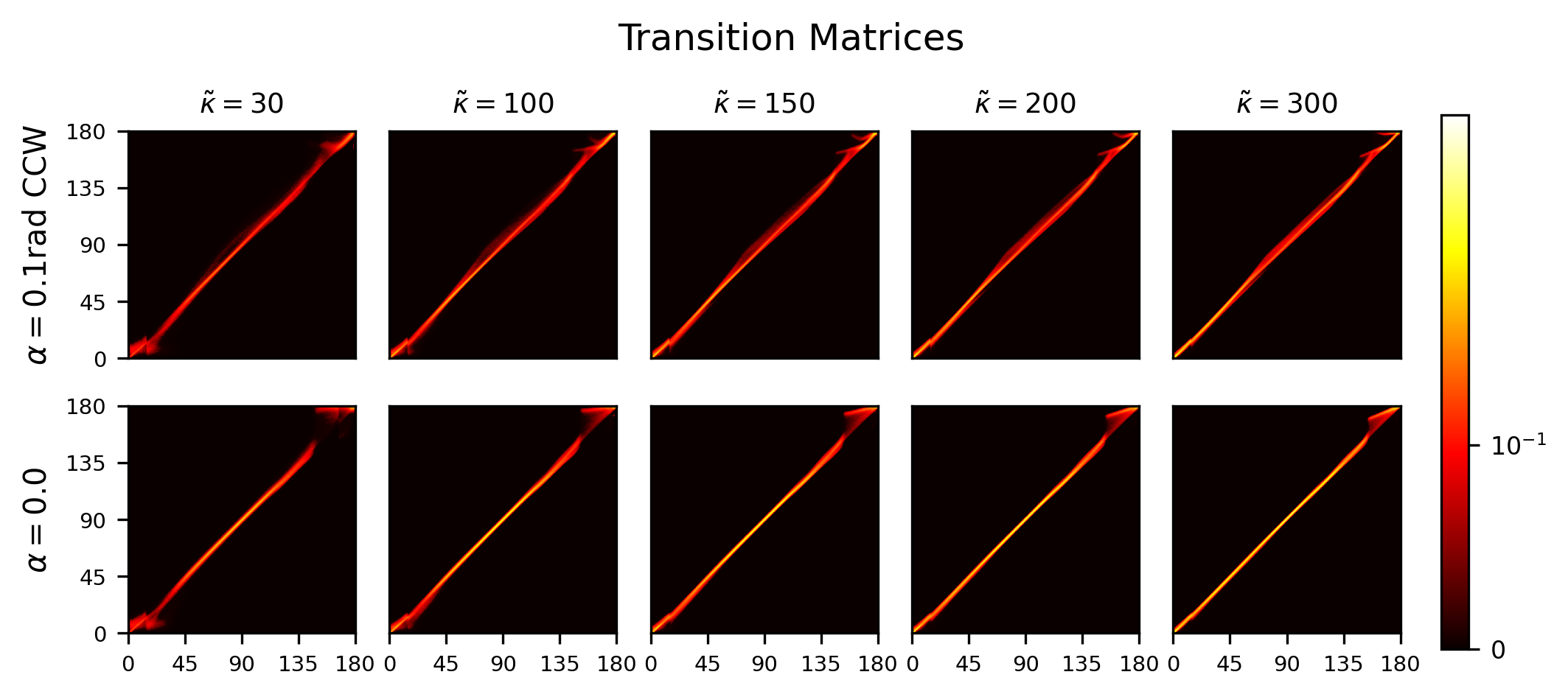}
    \caption[Transition matrix which represents the probability of having an incident angle of collision between two filaments $\psi_i$ and scattered angle as $\psi_j$.]{The transition matrix $T_{ij}$ which represents the probability of having an incident angle of collision between two filaments as $\psi_i$ (x-axis) and scattered angle as $\psi_j$ (y-axis). This is a heat map with hotter values indicating a higher probability of transitioning from $\psi_i \rightarrow \psi_j$.}
    \label{fig:transition_matrix}
\end{figure}

When a collision reorients filaments, it changes the population of filament orientations available for the next collision. As a bridge between our two-filament study and the collective behavior, we apply an iterative scheme to obtain predictions for the sign of the bulk rotation through $\Delta b$  (main text Fig.~3\textbf{E}).
%
Our iterative map treats the scattered angle $\psi_s$ as the incident angle $\psi_i$ in the next scattering event. For this purpose, we plot heatmaps in Fig.~\ref{fig:transition_matrix} of the probability distributions $T_{ij}$ for $\psi_s$ in bin $j$ at each $\psi_i$ in bin $i$; the mean and standard deviation of this data for each $\psi_i$ are the contents of Fig.~\ref{SI-binaryMT}\textbf{D,E}. We then interpret the transpose $T^T$ at each $\tilde \kappa$ as mapping a distribution of incident angles onto a distribution of scattered angles through matrix multiplication.

We consider the system's state to be represented by an occupancy column vector $v_{t}$ of incident angles, which are conceptually equivalent to the distribution of filament orientations about the mean order direction. Time evolution to timestep $t$ is then modeled as iterative multiplication $(T^T)^t v_{t_0}$ of the transpose of the transition matrix on the initial occupancy vector.
A uniform initial distribution $v_{t_0}$ represents an isotropic initial state. 

We seek to determine whether this system generically evolves toward a well-defined steady state.  We do not include random fluctuations in our simulated pair collisions, so each collision is deterministic and the scattering process applied to a distribution is Markovian \cite{Seabrook2023}. From the spectral theory of Markov chains \cite{Seabrook2023}, iterative multiplication of the transition matrix ($(T^T)^t$) will lead the system to converge to the matrix's stationary distribution, dependent on its eigenvalues. The largest eigenvalue $\lambda_0$ satisfies $\lambda_0\leq 1$, and can equal one only in trivial scenarios, due to the probabilistically normalized construction of $T$. We can list the eigenvalues in decreasing order as $\lambda_0 \ge \lambda_1 \ge ... \ge \lambda_d$. If the spectral value decomposition of $T^T$ is full rank, \textit{i.e.} if $T^T$ can be represented non-degeneratively in the basis of its eigenvectors $e_i$ (which is the case for our data), then we may rotate our occupancy vectors into the eigenbasis of the transition matrix $P$ by
\begin{equation}
    v=\sum_{i=0}^d \langle v,e_i \rangle e_i
\end{equation}
where $\langle v,e_i \rangle$ represents the projection of $v$ on $e_i$. By the linearity of the operator $T^T$, the time evolution acts independently on each eigenvector:
\begin{equation}
    (T^T)v=\sum_i \langle v,e_i \rangle ((T^T) e_i )=\sum_i \langle v,e_i \rangle \lambda_i e_i.
\end{equation}
More generally, after $t$ timesteps,
\begin{equation}
    (T^T)^t v = \sum_i \langle v,e_i \rangle (\lambda_i)^t e_i. \label{eq:iterative-time-evolution}
\end{equation}
By the restrictions on $\lambda_i$ to lie in the range $[0,1)$ in all nontrivial cases (excluding $\lambda_0=1$), we are guaranteed that  $(\lambda_i)^t$ decreases with increasing $t$. In the long-time limit, a generic distribution evolving according to Eq.~\ref{eq:iterative-time-evolution} will be dominated by the eigenvector $e_0$ corresponding to the largest eigenvalue of $T$. 

To approximate the long-time limit, for each sampled filament flexibility $\tilde{\kappa}$ and skew angle $\alpha$, we numerically find the number of iterations $t^*$ necessary to reach the stationary distribution $S = (T^T)^{t^*}$ such that 
\begin{equation}
|\langle (T^T)^{t^*} v_{t_0}, e_0 \rangle - \langle (T^T)^{t^*-1} v_{t_0}, e_0 \rangle | < 10^{-8}
\end{equation}
where $10^{-8}$ is an arbitrarily chosen small tolerance. 

\begin{figure*}[h!]
    \centering
    \includegraphics[width=17cm]{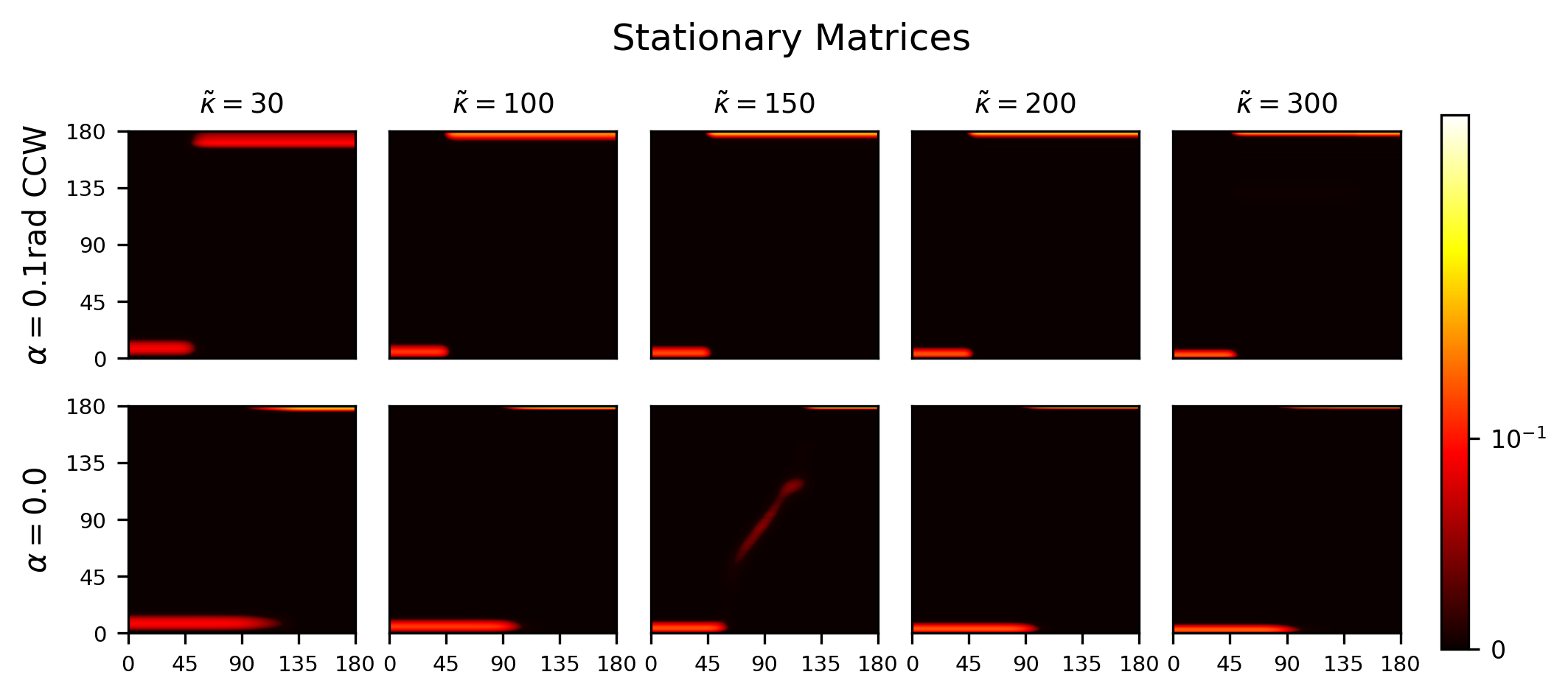}
    \caption[Stationary matrices obtained by iterative multiplication of the transition matrix]{Stationary matrices obtained by iterative multiplication of the transition matrix $S = (T^T)^{t^*}$ for $t^*$ computed numerically. $\alpha=0.1$ tends to favor LRO nematic steady states while $\alpha=0.0$ tends to favor LRO polar steady state in the bulk. The x-axis represents the incident angles and y-axis is the scattered angles.}
    \label{fig:stationay_matrix}
\end{figure*}

The resulting stationary matrices, plotted as heatmaps in Fig.~\ref{fig:stationay_matrix}, reveal important differences between the chiral and achiral systems. With $\alpha = 0$, the probability of occupancy is larger in the lower-angle regimes ($\psi_s<90^\circ$) which corresponds to a LRO polar steady state, in agreement with the bulk steady state seen in Ref.~\cite{athani2023symmetry}.  In contrast, when When chirality is introduced with $\alpha =0.1$ rad CCW, the probability of occupancy near $\psi_s = 180^\circ$ becomes much stronger, a signal of the antiparallel as well as parallel alignments characteristic of nematic order, as we observe in our chiral bulk simulations. We also note interesting behavior at $\tilde{\kappa}=150$  with $\alpha=0$, where there is a significant probability of intermediate scattering angles, representing non-negligible steady-state disorder despite the polar and nematic basins. We believe this disorder to correspond with the fraction of filaments persistently oriented at angles far from the direction of nematic order, which were much more noticeable in the achiral results of Ref.~\cite{athani2023symmetry} than in our chiral simulations here.

\begin{figure*}[h!]
    \centering
    \includegraphics[width=17cm]{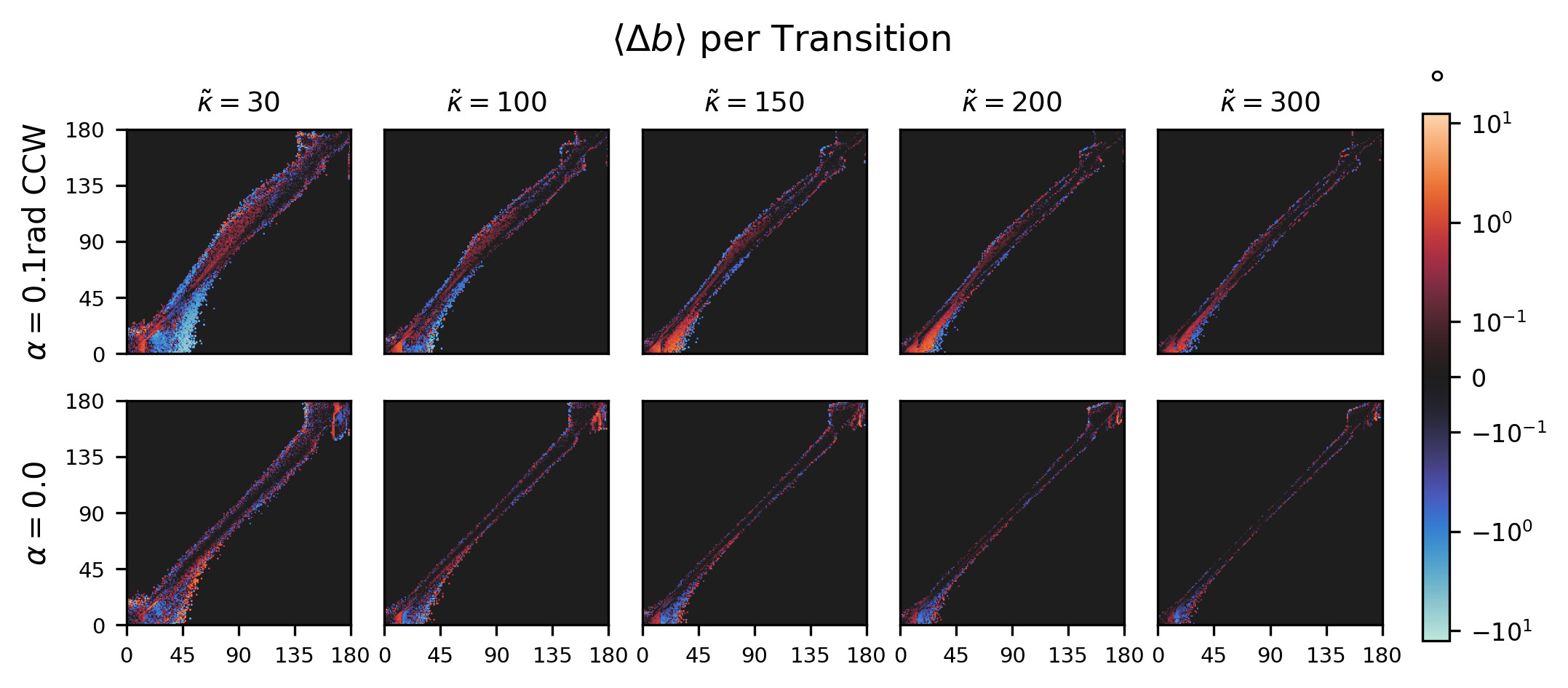}
    \caption[Average change in bisector per scattering transition.]{ Average change in bisector per scattering transition. The x-axis represents the incident angles and y-axis is the scattered angles. Colors indicate the amount of bisector rotation with negative values indicating CW rotation and positive values indicating CCW rotation. }
    \label{fig:bisector_per_transition}
\end{figure*}

To generate predictions for the rotational contribution of each collision, we calculate from our pair collision data the average bisector rotation $\langle \Delta b_{ij} \rangle$ associated with an incident angle in bin $i$  and a measured scattered angle in bin $j$. Heatmaps of this quantity are shown in Fig.~\ref{fig:bisector_per_transition}. Considering first the case of $\tilde{\kappa}=30$ and $\alpha =0.1$ rad CCW, we notice substantial breaking of mirror symmetry at near-parallel collision angles ($\psi_i<90^\circ$): there is more net CW rotation (blue colors) than CCW (red colors). Importantly, the predominant handedness of pair rotation is opposite in handedness to the CCW chiral activity, as we saw in the bulk for flexible filaments (main text Fig.~3\textbf{A},\textbf{C}). As we increase $\tilde{\kappa}$ to $100$ and then $150$, we see that the predominant handedness of pair rotation at small $\psi_i$ changes gradually from CW to CCW, in qualitative agreement with the bulk tendency toward CCW collective rotation at higher $\tilde \kappa$.
In contrast, the achiral case with $\alpha = 0.0$ has no systematic dominance of either handedness, as we verify quantitatively in the following analysis. 

\begin{figure*}[h!]
    \centering
    \includegraphics[width=\linewidth]{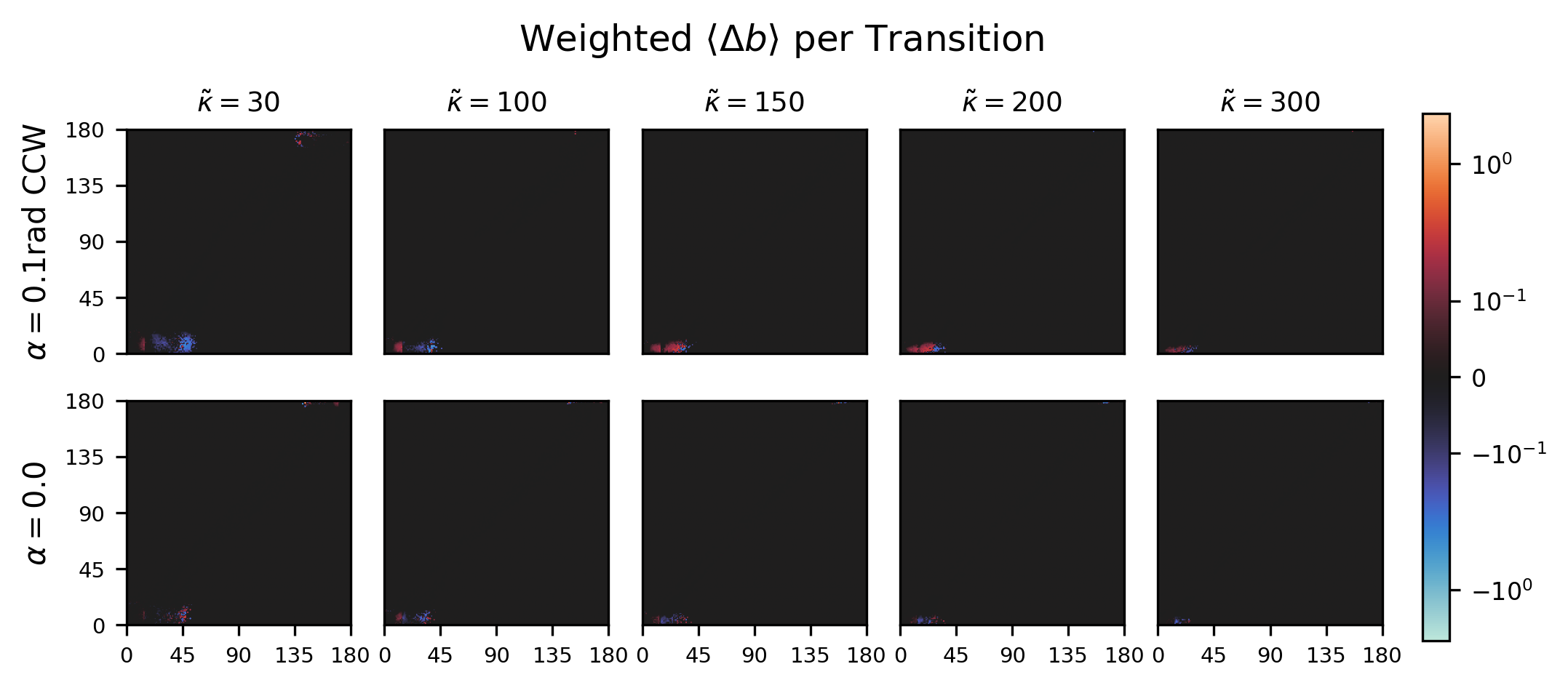}
    \caption[Average bisector rotation weighted by the stationary matrix.]{Average bisector rotation in Fig.~\ref{fig:bisector_per_transition} weighted by the stationary matrix from Fig.~\ref{fig:stationay_matrix} to calculate the rotation from each scattering event weighted by the estimated probability of that scattering event occurring in the steady state.}
    \label{Weight_deltab}
\end{figure*}

We now seek to transform our pair rotation calculations into a prediction for collective rotation in the bulk. We therefore weight the $\left<\Delta b\right>$ data for each pair of incident and scattered angles (Fig.~\ref{fig:bisector_per_transition}) by the probability of that scattering event occurring, estimated through the stationary matrix  $S_{ij}$ representing steady-state collision angle statistics. We therefore multiply each element of the matrix $\left<b_{ij}\right>$ by the corresponding entry in $S_{ij}$ (Fig.~\ref{fig:stationay_matrix}, and plot the results as a heatmap in Fig.~\ref{Weight_deltab}. Comparing Fig.~\ref{Weight_deltab} to Fig.~\ref{fig:bisector_per_transition}, we see that many types of collision with large unweighted $|\left<\Delta b\right>|$ contribute negligibly to the prediction bulk rotation due to the low expected rate of occurrence of such collisions. 
We operate this weighted $\left<\Delta b_{ij}\right>$  matrix on the initial  angle occupancy vector $v_{t_0}$, taken to be uniform, to produce a predicted distribution of rotation $\Delta b$  per collision in the bulk. Finally, we take the mean and standard deviation of each such distribution and plot this as $\left<\Delta b\right>$, with error bars, in Fig.~3\textbf{E} of the main text.

As seen in Fig.~3\textbf{E} of the main text, achiral filaments with $\alpha=0.0$ are predicted to exhibit no chiral bias in their rotation, as expected. The $\alpha=0.1$ rad CCW system rotates CW for $\tilde{\kappa}$ below a critical value $\tilde{\kappa}_c \approx 125$, and rotates CCW for larger $\tilde{\kappa}$. This prediction, based solely on pair collision data, qualitatively matches the change in handedness of collective rotation that we observed in bulk simulations upon increasing $\tilde{\kappa}$, with $\alpha$ held fixed. Quantitatively, the $\tilde \kappa_c$ predicted by the iterative scheme is smaller than that observed in bulk, but by a factor less than $2$, which we consider reasonable in light of the many approximations required in the above calculation. 


\begin{figure*}[h!]
    \centering
    \includegraphics[width=17cm]{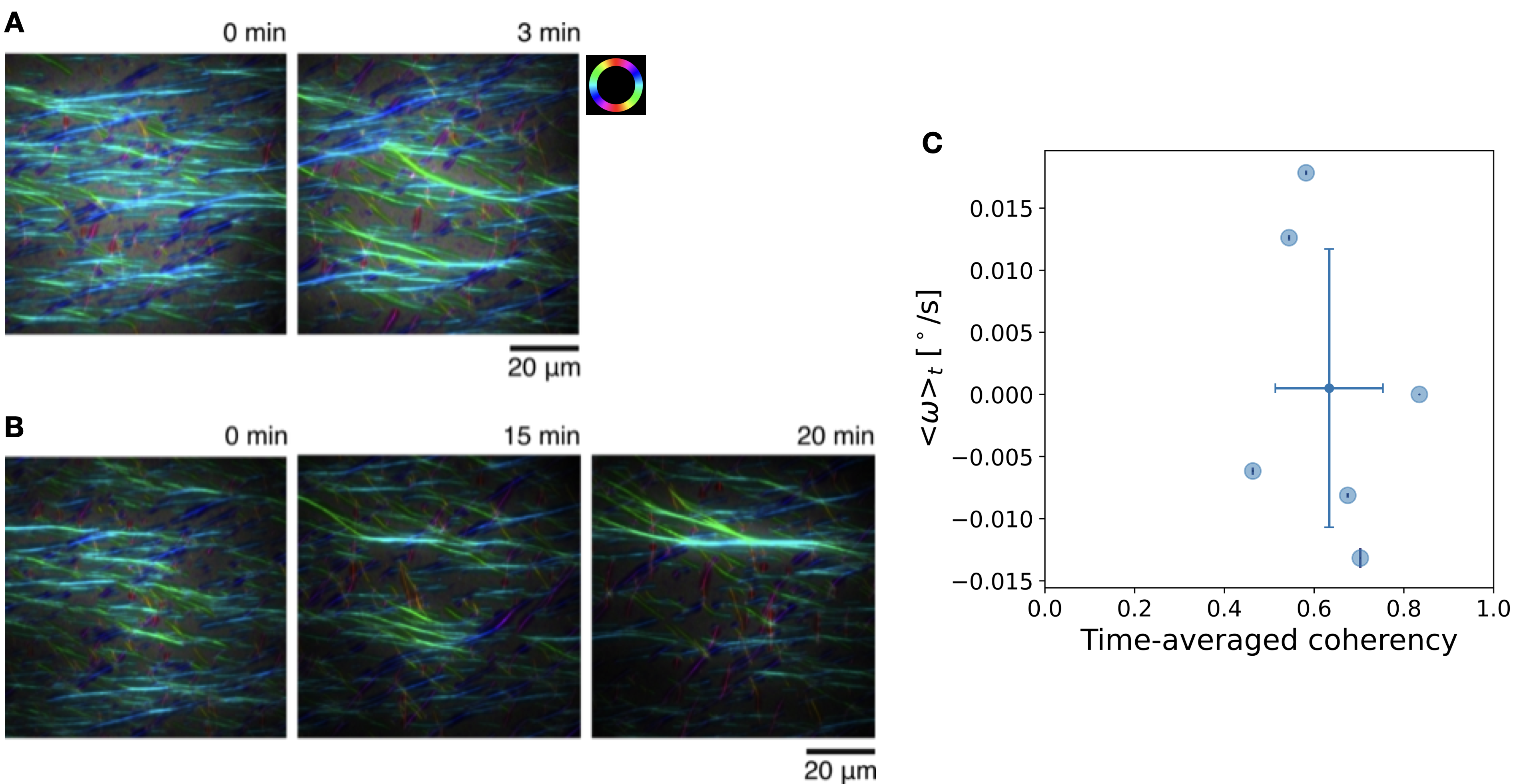}
    \caption{\textbf{A}, \textbf{B.} Snapshots of experiments with taxol-stabilized MTs on a glass substrate. \textbf{C.} Relation between the coherency of the nematic order and the rotation rate of the nematic director. Data from individual experiments are represented as error bars with median central line and whiskers extending to $\pm$ s.d.}
    \label{SI-taxol_glass}
\end{figure*}

\begin{figure*}[h!]
    \centering
    \includegraphics[width=17cm]{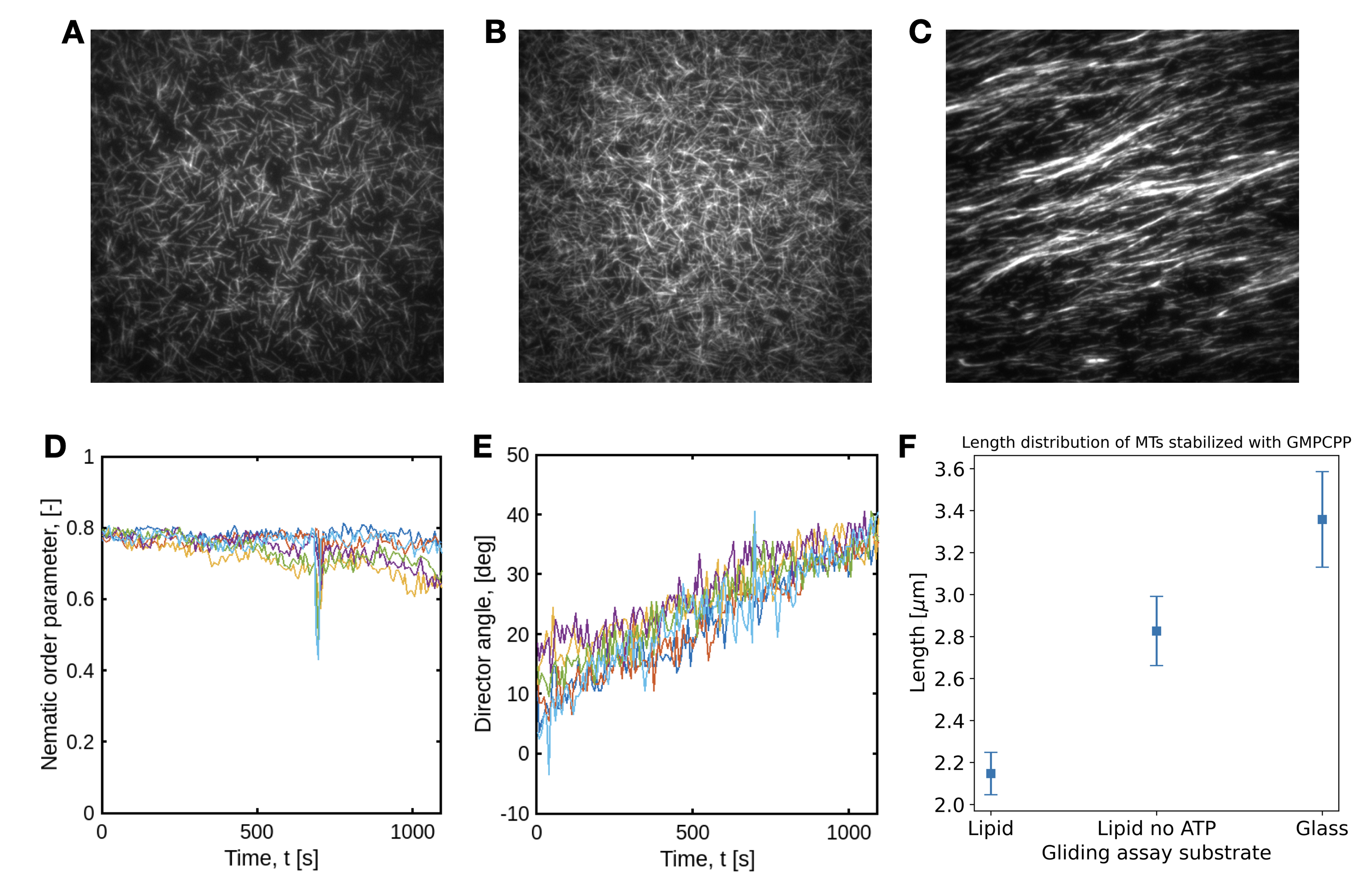}
    \caption{The influence of motor activity on the emergence of active nematic organization on lipid substrates. \textbf{A.} MTs imaged right after their adsorption to the lipid surface (no ATP). \textbf{B.} MTs imaged after the washing of the channel (no ATP). This shows that the hydrodynamic ordering is not responsible for the emergence of the nematic order. \textbf{C.} MTs imaged after the addition of ATP. The order emerges in less than a minute. \textbf{D.} Nematic order parameter as a function of time. \textbf{E.} Orientation of the nematic director with time. \textbf{F.} Average microtubule lengths with different gliding assay substrates. The error bars represent standard error of the mean. 41 filaments were chosen randomly from dense gliding assay systems.}
    \label{SI-GMPCPP_lipid}
\end{figure*}

\begin{figure*}[h!]
    \centering
    \includegraphics[width=17cm]{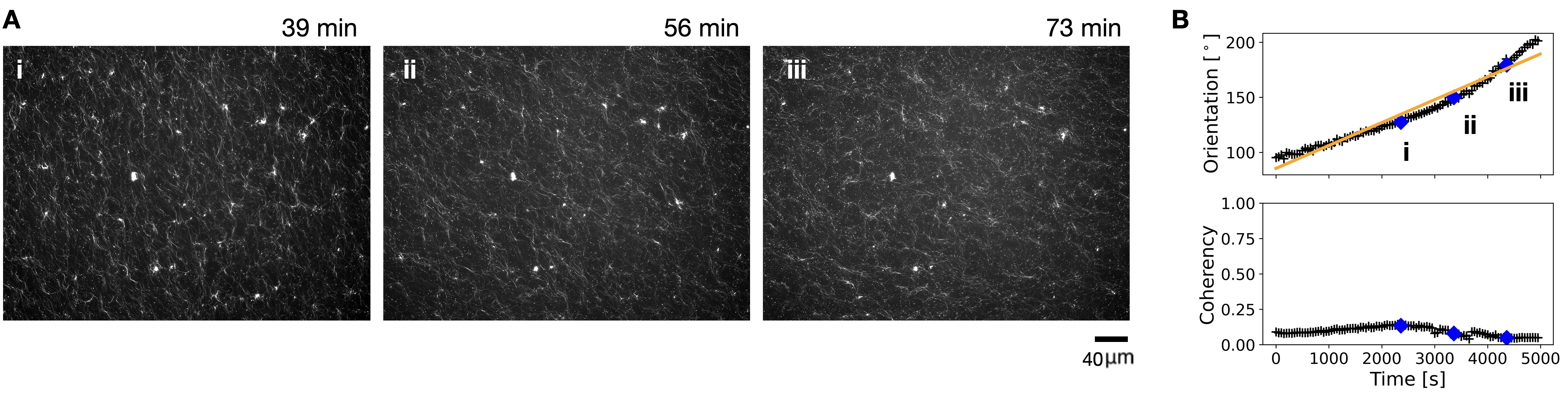}
    \caption{\textbf{A.} Snapshots from a experiment with taxol-stabilized MTs on lipid substrates. The field of view here is $\approx$450 x 335 $\mu$m. Each panel is an overlay of 10 consecutive frames (total time = 100 s). The starting time is marked on each panel. \textbf{B.} The temporal profile of the nematic director orientation for the data in \textbf{A} where the entire field of view is seen to rotate. Since this is a large field of view compared to the other cases, we take patches of this data and plot it in Fig.~\ref{SI-taxol_lipid_patch}.}
    \label{SI-taxol_lipid_full}
\end{figure*}

\begin{figure*}[h!]
    \centering
    \includegraphics[width=17cm]{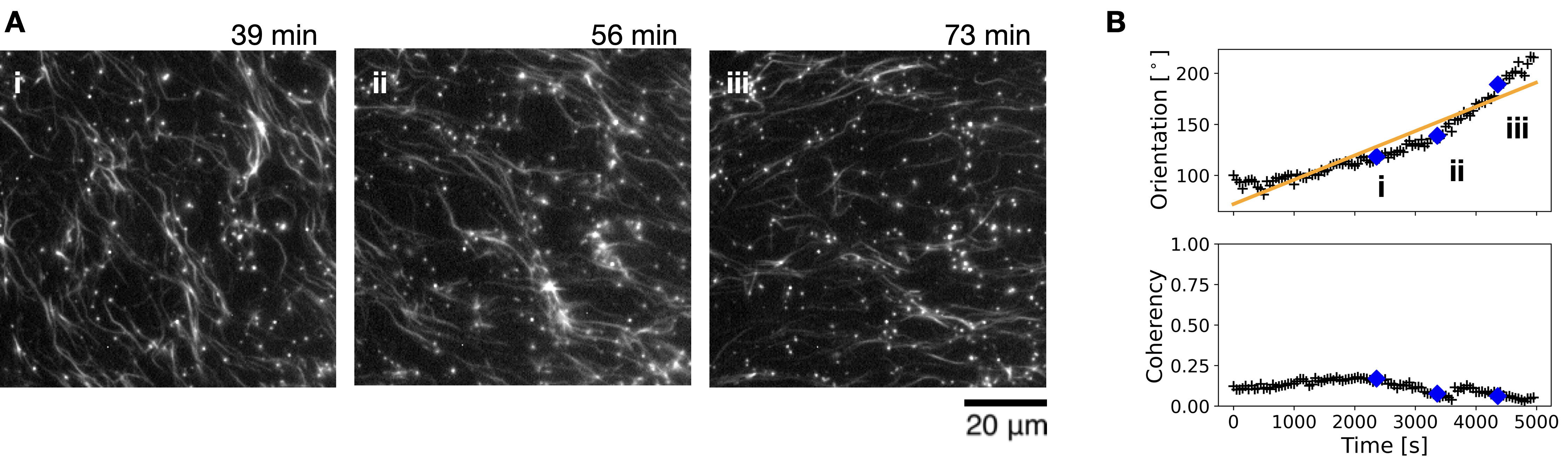}
    \caption{\textbf{A.} Snapshots from a experimental patch from Fig.~\ref{SI-taxol_lipid_full} with a $\approx 81 \times 81 \mathrm{\mu m}$ field of view. Each panel is an overlay of 10 consecutive frames (total time = 100 s). The starting time is marked on each panel. \textbf{B.} The temporal profile of the nematic director orientation for the data in \textbf{A} where the positive slope represents CCW rotation.}
    \label{SI-taxol_lipid_patch}
\end{figure*}

\section{Fragmented Lipid}

\begin{figure*}[h!]
    \centering
    \includegraphics[width=17cm]{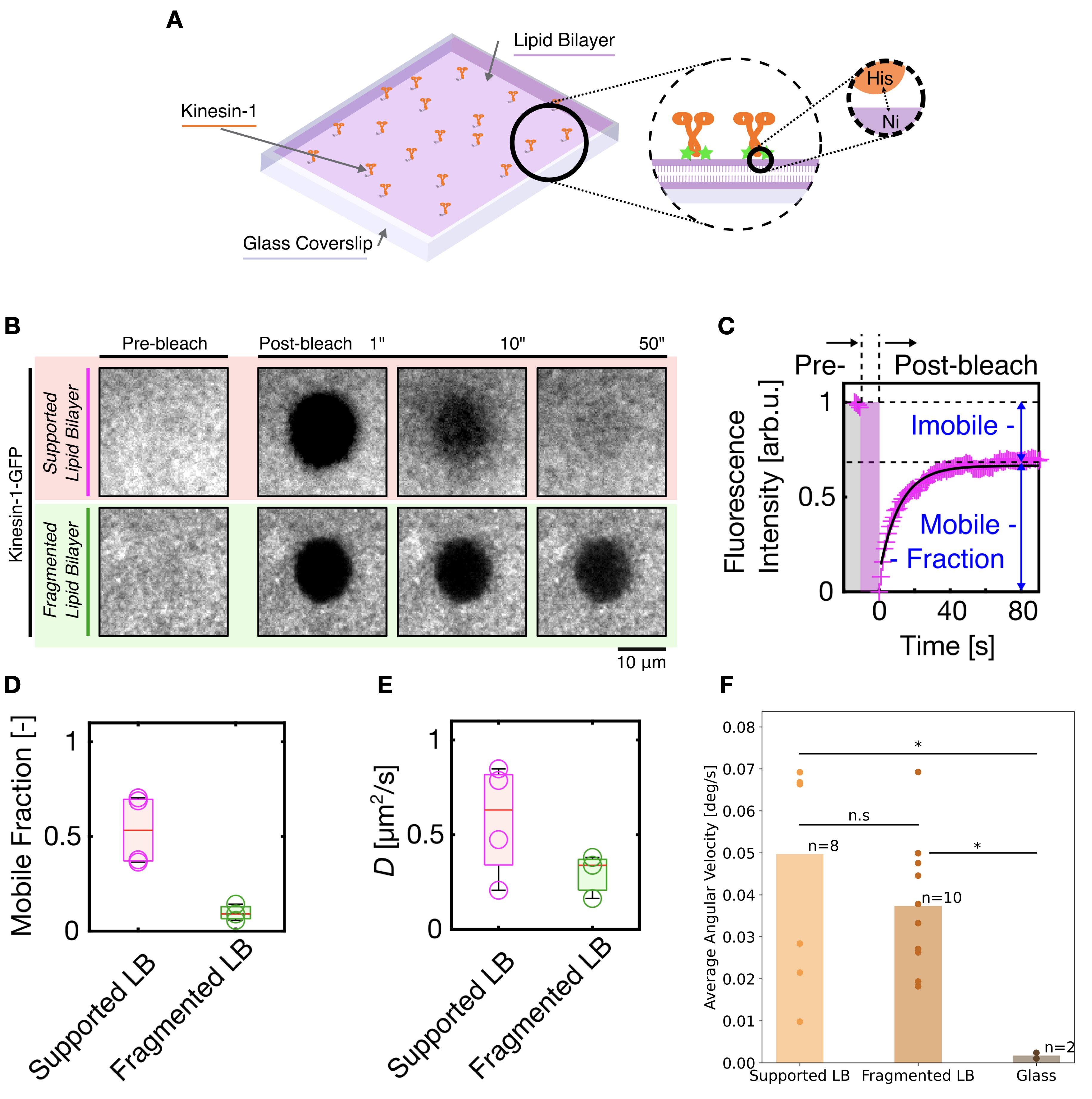}
    \caption{\textbf{A.} Schematic representation of the anchorage of kinesin-1 motors to the lipid bilayer via histidine tag – nickel link. \textbf{B.} Fluorescence recovery after photobleaching (FRAP) of the kinesin motors in the bilayer. \textbf{C.} A typical example of the average fluorescence signal in the photo-bleached area in the pre-bleach and post-bleach phases of the FRAP experiment. \textbf{D.} Mobile fraction of kinesin motors is substantially reduced in the fragmented lipid bilayer compared to the intact supported lipid bilayer. \textbf{E.} The diffusion constant of the motors. \textbf{F.} Rotation rates of the director of the LRO nematic formed on there different substrates.}
    \label{SI-fragmented_lipid}
\end{figure*}

Motors attached to lipid bilayers produce reduced force due to their lateral diffusion within the membrane, resulting in slipping \cite{grover2016transport}. We hypothesized that restricting the lateral mobility of motors by forming a fragmented lipid bilayer—where motor proteins are confined within smaller, isolated membrane regions—would create conditions intermediate between lipid-based assays and assays performed on glass. To evaluate this hypothesis, we prepared fragmented lipid bilayers and conducted gliding assays using GMPCPP-microtubules (Methods). While the ordering parameter observed in this fragmented system was somewhat lower {
(0.46$\pm$0.04, n = 10 experiments)}, the cholesteric rotation rate was markedly reduced, averaging {
0.037 $\pm$ 0.005 degrees} per second (\ref{SI-fragmented_lipid}).

\begin{figure*}[h!]
    \centering
    \includegraphics[width=17cm]{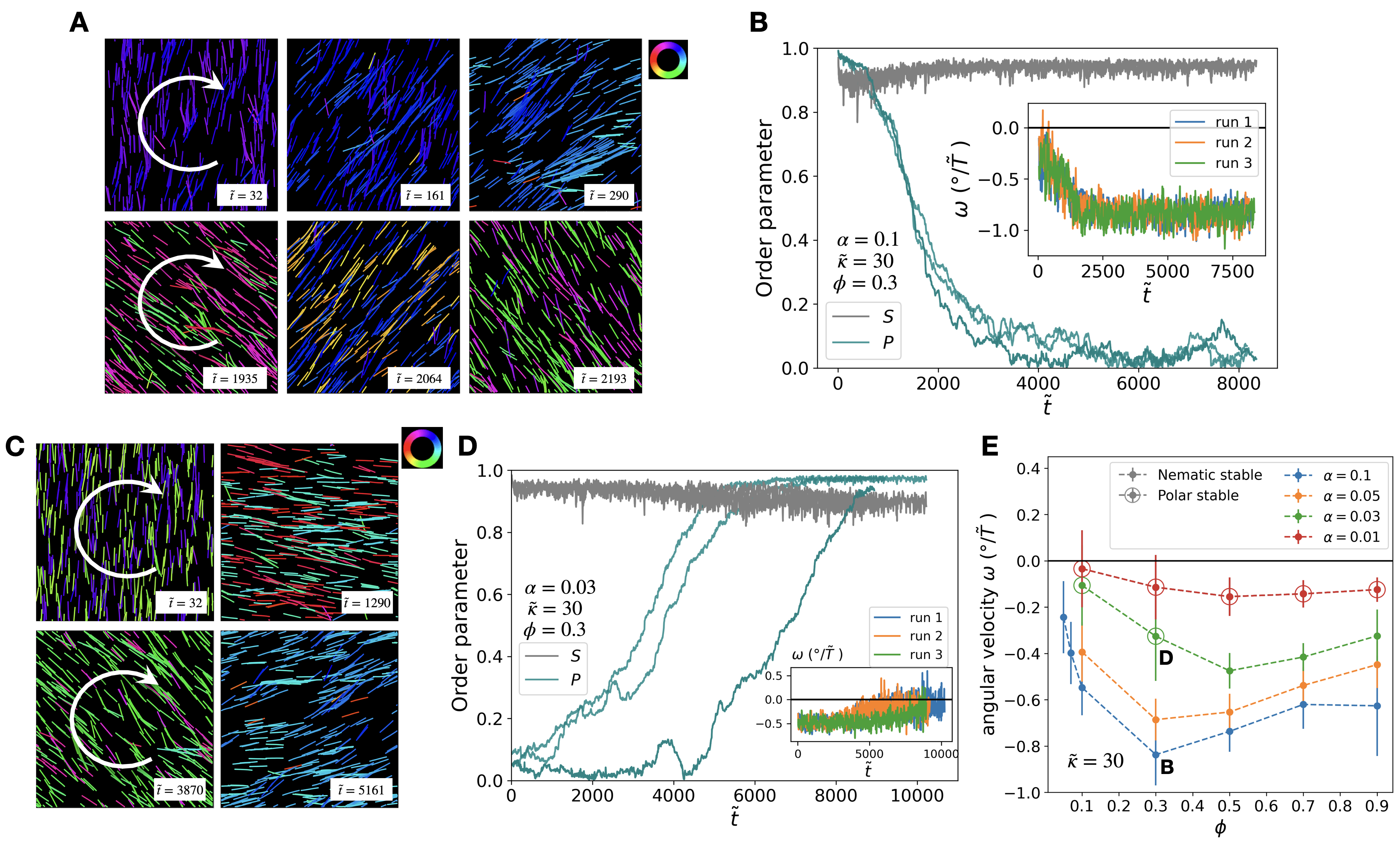}
    \caption{\textbf{A, B.} Evolution from a polar initialization toward a nematic steady state; $\alpha = 0.1$rad, $\tilde \kappa = 30$, $\phi= 0.3$, $L_x = 150$. \textbf{A.} (Same as main text Fig.~5\textbf{A}) Snapshots of time evolution in a typical simulation. \textbf{B.} Time-series plots of the orientational order parameters, overlaying the data from three independent runs. The polar order parameter $P$ decays while the nematic order parameter $S$ remains high. Inset shows the angular velocity, which saturates at the same negative value in each run. \textbf{C, D.} Evolution from a nematic initialization toward a polar steady state; $\alpha = 0.03$rad, $\tilde \kappa = 30$, $\phi= 0.3$, $L_x = 150$. \textbf{C}. Snapshots of time evolution in a typical simulation. \textbf{D.} Time-series plots of the orientational order parameters, overlaying the data from three independent runs. The polar order parameter rises to nearly $1$ while the nematic order parameter remains high. Inset shows the angular velocity, which is generally negative but decreases in magnitude. \textbf{E.} Steady-state angular velocity as a function of area fraction $\phi$ and counterclockwise skew angle $\alpha$, with fixed bending rigidity $\tilde \kappa = 30$. Each data point shows the mean from a 3-run ensemble; error bars show standard deviation. 
    Circled points indicate an LRO polar steady state; the others are LRO nematic. 
    The angular velocity decreases in magnitude as $\alpha$ is decreased from  $0.1 \rightarrow 0.01\mathrm{rad}$ CCW, for each value of $\phi$. The angular velocity also decreases in magnitude as $\phi$ is reduced from  $0.3 \rightarrow 0.05$ keeping $\alpha = 0.1$rad CCW fixed. No LRO state was obtained with $\phi < 0.05$ for $\alpha = 0.1\mathrm{rad}$. The data points corresponding to B, D are labeled.
    }
    \label{SI-changing_alpha}
\end{figure*}

\begin{figure*}[h!]
    \centering
    \includegraphics[width=15cm]{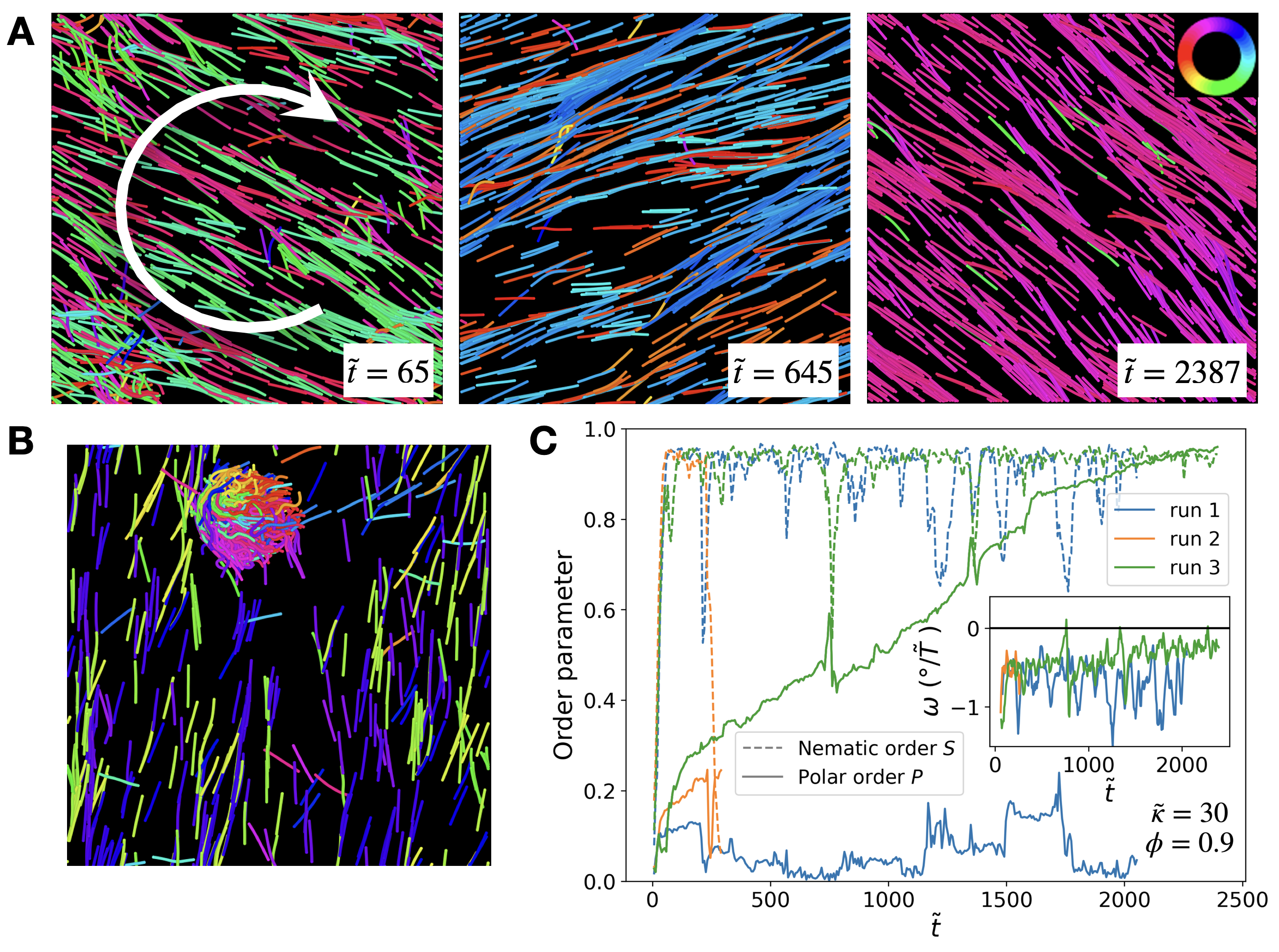}
    \caption{Plots showcasing the 'multiple outcomes' category of the phase diagram. \textbf{A.} (Same as main text Fig.~5\textbf{D}) Snapshots of a simulation whose director rotates clockwise with LRO polar order as its steady state; $\tilde{\kappa} = 30$, $\phi = 0.9$. \textbf{B.} (Same as main text Fig.~5\textbf{E}) Example of a motile cluster formed within a previously uniform LRO nematic. Motile clusters can either be transient or can grow to eventually consume all filaments. \textbf{C.} Time-evolution of polar (solid lines) and nematic (dashed lines) order parameters plotted for three different simulation cases: Polar stable (green; corresponding to \textbf{A}), Nematic stable (blue), and motile clusters (orange; corresponding to \textbf{B}); the latter results in early termination of the simulation run. Inset shows time-series of the angular velocity for all three cases.
     }
    \label{SI-multiple_outcomes}
\end{figure*}

\section{List of Supplementary Movies}

\begin{itemize}
\item \textit{\textbf{Movie S1: Rotating active nematic on glass substrate, higher microtubule density}} \\ 
Fluorescence micrograph time-series of a rotating active nematic state in a gliding assay of GMPCPP-microtubules on a glass substrate, with microtubule surface density $0.20 \pm 0.02$, corresponding to snapshots in Fig.~1\textbf{A}. 

\item \textit{\textbf{Movie S2: Rotating active nematic on glass substrate, lower microtubule density}} \\ 
Fluorescence micrograph time-series of a rotating active nematic state in a gliding assay of GMPCPP-microtubules on a glass substrate, with microtubule surface density $0.13 \pm 0.01$, corresponding to snapshots in Fig.~1\textbf{B}. 
%
\item \textit{\textbf{Movie S3: CCW rotation in simulation with higher filament rigidity}} \\
Simulation exhibiting CCW rotation of the nematic director for bending rigidity $\tilde{\kappa} = 300$, skew angle $\alpha = 0.1$ rad CCW, and area fraction $\phi = 0.3$, corresponding to snapshots in Fig.~2\textbf{C}. Filaments are colored by the average tangent orientation of each bead in the bead-spring chain (color wheel legend). The framerate corresponds to $36.5\pm 0.8$ $\tilde T$ per second, where $\tilde T = l / v_0$ is the active timescale  and the small variability is due to adaptive time-stepping.

\item \textit{\textbf{Movie S4: No rotation in simulation with intermediate filament rigidity}} \\
Simulation exhibiting no significant rotation of the nematic director for bending rigidity $\tilde{\kappa} = 200$, skew angle $\alpha = 0.1$rad CCW, and area fraction $\phi = 0.3$, corresponding to snapshots in Fig.~3\textbf{A} top row. Filaments are colored by the average tangent orientation of each bead in the bead-spring chain (color wheel legend). The framerate corresponds to $52.1\pm 1.0$ $\tilde T$ per second, where $\tilde T = l / v_0$ is the active timescale  and the small variability is due to adaptive time-stepping.

\item \textit{\textbf{Movie S5: CW rotation in simulation with lower filament rigidity}} \\
Simulation exhibiting CW rotation of the nematic director for bending rigidity $\tilde{\kappa} = 100$, skew angle $\alpha = 0.1$ rad CCW, and area fraction $\phi = 0.3$, corresponding to snapshots in Fig.~3\textbf{A} bottom row. Filaments are colored by the average tangent orientation of each bead in the bead-spring chain (color wheel legend). The framerate corresponds to $102.5\pm 2.9$ $\tilde T$ per second, where $\tilde T = l / v_0$ is the active timescale  and the small variability is due to adaptive time-stepping.

\item \textit{\textbf{Movie S6: Rotating active nematic on lipid bilayer}} \\
Fluorescence micrograph time-series of the GMPCPP-microtubule gliding assay on lipid bilayer, exhibiting an active nematic state with coherent CCW rotation of the director, corresponding to snapshots in Fig.~4\textbf{A}.
    %

\item \textit{\textbf{Movie S7: Multiple outcomes in simulation with high filament density}} \\
Simulation in the \textit{Multiple outcomes} regime of high density ($\phi = 0.9$) and low bending rigidity ($\tilde \kappa = 30$) on the active phase diagram of Fig.~5\textbf{F}. The active force skew angle is $\alpha = 0.1$ rad CCW. The system is initialized isotropically. Transient active states exhibited include active nematic with CW rotation (snapshots in Fig.~5\textbf{D} left and middle panels), polar bands traveling transversely, and clusters (see Fig.~5\textbf{E} for a snapshot of a stable cluster in a different simulation with the same parameters). The steady state in this run is apparently active polar with CW rotation (snapshot in Fig.~5\textbf{D} right panel). The framerate corresponds to $166.6\pm 71.0$ $\tilde T$ per second, where $\tilde T = l / v_0$ is the active timescale  and the variability is due to adaptive time-stepping.
\end{itemize}

\end{document}